\documentclass[prb,twocolumn, showpacs,amsmath,amssymb,superscriptaddress,floatfix]{revtex4-1}
\usepackage{amsmath}
\usepackage{amssymb}
\usepackage{amsthm}
\usepackage{amsfonts}
\usepackage{dsfont}
\usepackage{xcolor}
\usepackage{listings}
\usepackage{enumerate}
\usepackage{latexsym}
\usepackage{mathdots}

\usepackage{psfrag}

\usepackage{color}

\usepackage{bm}
\usepackage{graphicx}
\usepackage{subfigure}

\newcommand{\mm}{\text{-}1}

\newcommand{\spa}{\text{ }}

\newcommand{\beq}{\begin{equation}}
\newcommand{\eneq}{\end{equation}}

\newcommand{\braket}[2]{\left\langle #1 | #2 \right\rangle}
\newcommand{\bra}[1]{\left\langle#1\right|}
\newcommand{\ket}[1]{\left|#1\right\rangle}

\DeclareMathOperator{\sech}{sech}
\DeclareMathOperator{\arcsech}{arcsech}
\DeclareMathOperator{\arcsec}{arcsec}

\newcommand{\mJ}{\mathcal{J}}
\newcommand{\bM}{\boldsymbol{M}}

\input{epsf}

\newcommand{\sumal}[2]{\underset{#1}{\overset{#2}{\sum}}}

\newcommand{\nn}{\nonumber}

\newcommand{\twopartdeft}[3]
{
	\left\{
		\begin{array}{ll}
			#1 & \mbox{\textrm{if} } #2 \\
			#3 & \mbox{\textrm{otherwise} }
		\end{array}
	\right.
}

\newcommand{\twopartdef}[4]
{
	\left\{
		\begin{array}{ll}
			#1 & \mbox{\textrm{if} } #2 \\
			#3 & \mbox{\textrm{if} } #4
		\end{array}
	\right.
}

\newcommand{\threepartdef}[6]
{
	\left\{
		\begin{array}{ll}
			#1 & \mbox{\textrm{if} } #2 \\
			#3 & \mbox{\textrm{if} } #4 \\
			#5 & \mbox{\textrm{if} } #6
		\end{array}
	\right.
}

\begin{document}

\tolerance 10000

\newcommand{\vk}{{\bf k}}


\title{Operator Spreading in Quantum Maps}

\author{Sanjay Moudgalya}
\affiliation{Department of Physics, Princeton University, Princeton, NJ 08544, USA}
\author{Trithep Devakul}
\affiliation{Department of Physics, Princeton University, Princeton, NJ 08544, USA}
\author{C. W. von Keyserlingk}
\affiliation{University of Birmingham, School of Physics \& Astronomy, B15 2TT, UK}
\author{S. L. Sondhi}
\affiliation{Department of Physics, Princeton University, Princeton, NJ 08544, USA}
\date{\today}

\begin{abstract}
Operators in ergodic spin-chains are found to grow according to hydrodynamical equations of motion.
The study of such operator spreading has aided our understanding of many-body quantum chaos in spin-chains. 
Here we initiate the study of ``operator spreading" in quantum maps on a torus, systems which do not have a tensor-product Hilbert space or a notion of spatial locality. 
Using the perturbed Arnold cat map as an example, we analytically compare and contrast the evolutions of functions on classical phase space and quantum operator evolutions, and identify distinct timescales that characterize the dynamics of operators in quantum chaotic maps. 
Until an Ehrenfest time, the quantum system exhibits \emph{classical chaos}, i.e. it mimics the behavior of the corresponding classical system.  
After an operator scrambling time, the operator looks ``random" in the initial basis, a characteristic feature of \emph{quantum chaos}.
These timescales can be related to the quasi-energy spectrum of the unitary via the spectral form factor. 
Furthermore, we show examples of ``emergent classicality" in quantum problems far away from the classical limit.
Finally, we study operator evolution in non-chaotic and mixed quantum maps using the Chirikov standard map as an example. 
\end{abstract}

\maketitle

\section{Introduction}
The study of chaos and ergodicity in many body systems has recently acquired a major revival of interest, with the discovery of phenomena such as many-body localization,\cite{nandkishore2015many, huse2013localization, pal2010many, kjall2014many} and its connections to several fundamental questions regarding black holes, the scrambling of quantum information and quantum gravity.\cite{maldacena2016bound, cotler2017black, maldacena2016remarks, kitaev2015simple} 
This has led to the many recent explorations of the dynamics of quantum systems by means of diagnostics that are sensitve to such questions, many of which build
on advances in quantum information theory.
For example, quantum chaos has been explored both analytically and numerically in several systems by means of operator and entanglement growth,\cite{nahum2017quantum, nahum2017operator, von2017operator, mezei2017entanglement1, mezei2017entanglement, rakovszky2017diffusive, khemani2017operator, jonay2018coarse} behavior of Out-of-Time Ordered Correlators (OTOC),\cite{maldacena2016bound, maldacena2016remarks, aleiner2016microscopic, swingle2016measuring, rozenbaum2017lyapunov, rozenbaum2018universal,  xu2018accessing, khemani2018velocity} random matrix theory,\cite{cotler2017black, d2016quantum, kos2017many, zirnbauer2015symmetries, mondaini2016eigenstate} and a variety of other methods.\cite{cotler2017chaos, roberts2017chaos, ho2017ergodicity, torres2017dynamical, torres2018generic, li2018quantum, schiulaz2018thouless}
These studies have led to the introduction of new physical quantities which have shed light on the definition and meaning of many body quantum chaos. These include the butterfly and entanglement velocities defined using operator and entanglement growths, and frame potentials defined using the concept of unitary designs from random matrix theory. 

Of particular interest to this paper is the use of the Heisenberg picture---in the analysis of operator evolution and in the calculation of OTOCs. This use of the Heisenberg picture, which is standard in the study of quantum field theory and many body systems, is relatively new to the analysis of quantum chaos. Here the traditional approach, 
which was primarily developed in the study of single particle systems \cite{stockmann2000quantum} relies, as single particle quantum mechanics does typically, on the
Schr\"odinger picture. Our overall aim in this paper is to re-examine single particle quantum chaos in the Heisenberg picture building on the insights generated in the study
of many body quantum chaos.
%

%
Before turning to this re-examination we briefly mention some landmarks in the study of single-particle quantum chaos which is by now a venerable and well-developed
subject.
The observation that the differences in quantizations of classically regular and chaotic systems is exhibited in the eigenstate level statistics has been known for long, and it led to the notion of quantum chaos. \cite{bohigas1984characterization, berry1977level}
Several examples of quantum chaos including quantizations of classical billiards, the quantum kicked rotor, and more generally quantum maps \cite{berry1979quantum, izrailev1990simple, stockmann1990quantum, kottos1997quantum, delande1986quantum, haake1987classical, chirikov1988quantum, schack2000shifts, bianucci2002decoherence, keating2006nodal} were subsequently studied. 
Apart from the study of the eigenstate level statistics, which is predicted by random matrix theory for chaotic quantum systems, several other diagnostics of quantum chaos were subsequently used (see for example Ref.~[\onlinecite{haake2013quantum}] for a review). 
These include certain semi-classical ``trace formulae" that relates classical orbits of the classical system to the density of states of the quantized system,\cite{bogomolny1996gutzwiller, gutzwiller1980classical} as well as quantum measures, for example spectral form factors,\cite{kottos1997quantum, heusler2004universal} nodal domains,\cite{blum2002nodal} as well as the well-known Loschmidt echo.\cite{jacquod2001golden, quan2006decay, cucchietti2003decoherence}
Several works have studied quantum chaos in the Schr\"odinger picture employing phase-space representations of wavefunctions such as the Wigner and Husimi functions.\cite{takahashi1985chaos, zurek1994decoherence, karkuszewski2002quantum}
Finally, some works have even explored the Heisenberg picture, \cite{haake1987classical, cohen1991quantum} although the time-evolved operators were not directly studied.

As noted above, in this work, we aim to fill this lacuna by studying quantum chaos in single-particle systems via the Heisenberg evolution of operators. Naively, the Heisenberg evolution resembles the evolution in phase space as the operator evolution equations are quantized versions of the classical dynamical equations. As the solutions of the latter exhibit classical chaos we are led to ask how we can diagnose chaos in a quantum 
system from studying the evolution of operators. The work on many body systems leads to more specific
questions, e.g. is there a notion of operator spreading in a quantum chaotic systems that do not have a tensor product Hilbert space and what does this say about quantum chaos?

To answer these, we focus on quantum maps on a toroidal phase space which we review in Sec.~\ref{sec:classicalmaps}, for example the Arnold cat map and its perturbed version, whose classical limits have been well-studied.\cite{lasota1985probabilistic, backer2003numerical, agam1995semiclassical, kurlberg2001quantum} 
Studying the evolution of operator coefficients in a fixed basis (analogous to Pauli strings in spin-chains\cite{von2017operator, khemani2017operator, rakovszky2017diffusive}) provides a natural single-particle parallel to the studies of operator spreading in many-body quantum systems.
We choose an operator basis that maps on to the Fourier basis of smooth functions on phase space in the classical limit, allowing us to compare and contrast quantum operator evolutions and their classical counterparts, smooth functions on phase space. 
We introduce the basis and derive the equations of motion for operators as well as classical functions in Sec.~\ref{sec:heisenbergpicture}.
Before we study quantum evolution, we define the \emph{classicality} of evolution from an operator point of view in Sec.~\ref{sec:cqevolution}.
There, we distinguish the usual semi-classical (coherent state) basis and the Fourier basis in which we study operator evolution, and demonstrate the existence of a kind of classicality that arises away from the usual classical limit of the quantum map, which we call \emph{emergent classicality}.
The evolution of operators in such systems are analogous to Clifford circuits, where operators do not spread in a certain basis. 
In Sec.~\ref{sec:regimes}, we study operator evolution in the quantum chaotic limit (defined by the usual diagnostics of quantum chaos), where we show that the set of non-vanishing operator coefficients ``spread" over the operator basis, diagnosed by the growth of Shannon entropy of the set of operator coefficients.
We identify three distinct regimes of operator evolution: (i) Early-time semi-classical evolution (up to an Ehrenfest time $t_E$), where the operator evolution in the quantum problem mimics the evolution of smooth functions on classical phase space, (ii) Intermediate-time evolution, where the operators start to deviate from the classical behavior and maximally spread in operator space, and (iii) Late-time evolution (after an operator scrambling time $t_{\textrm{scr}}$) where the operator evolution has no classical analogue and exhibits features of random matrix theory.
The regime (i) has been well-studied in works of classical-quantum correspondence where it is known that the quantum system behaves classically up to an ``Ehrenfest time" that depends on the Lyapunov exponent of the classical system and the Planck's constant. \cite{helmkamp1994structures, ballentine1994inadequacy, fox1994chaos}
On the other hand, regime (iii) has been studied in the context of thermalization of isolated systems, where late-time expectation values of observables are determined by the ``diagonal ensemble", the expectation values of observables in the eigenstates of the time evolution unitary.\cite{rigol2008thermalization, nandkishore2015many} 
We show the basis-independence of the results in Sec.~\ref{sec:spectrum} (up to certain caveats that we discuss), and connect operator evolution to the spectral form factor, a well-known diagnostic of quantum chaos.  
Finally in Sec.~\ref{sec:quantumscars}, we discuss operator evolution in the Chirikov Standard Map, an example of a quantum map that is not completely completely quantum chaotic. There we show that regular and chaotic regions in phase space can be distinguished using the operator evolution diagnostics we discuss.  
In all, we show the existence of ``operator spreading" that occurs in quantum maps which can be used to characterize various regimes of quantum operator evolution, and distinguish chaotic quantum maps from non-chaotic ones.

\section{Classical Maps and their Quantization}\label{sec:classicalmaps}
We first review the historical introduction of quantum maps by ``quantizing" maps defined on a classical phase space.
Note that the quantum map thus obtained from a given classical map is not necessarily unique.
In this work we choose the classical map obtained by a particular quantization prescription and view the quantum map as the fundamental system that exhibits a classical description in a certain limit.

\subsection{Classical Maps}
We start with area-preserving discrete-time classical maps on phase space that has the topology of a torus $\mathbb{T}^2$, defined by coordinates 
\begin{equation}
    0 \leq q < 1, \;\;\; 0 \leq p < 1.
\end{equation}
Under the action of the map, any point on the torus at time $t$,  $(q, p) \equiv (q(t), p(t))$ is mapped to another point on the torus $(q', p') \equiv (q(t+1), p(t+1))$, such that the Jacobian of the transformation is 1. 
A class of well-known area-preserving maps are the Arnold Cat Maps. \cite{arnold1968ergodic, benatti1991non} These read
\begin{equation}
    \begin{pmatrix}
        q' \\
        p'
    \end{pmatrix}
    =
    \begin{pmatrix}
        a & b \\
        c & d
    \end{pmatrix}
    \begin{pmatrix}
        q \\
        p
    \end{pmatrix}
    {\rm mod\spa 1}
\label{eq:gencatmap}
\end{equation}
where $a, b, c, d$ are non-negative integers satisfying the area-preserving condition $ad - bc = 1$.
The Lyapunov exponent of the cat map of Eq.~(\ref{eq:gencatmap})
%
%
is given by $\lambda = \log \left( (a + d + \sqrt{a^2 + 4 b c - 2 a d + d^2})/2\right)$.
If $a + d > 2$, the cat map is chaotic, i.e. it has a non-zero Lyapunov exponent. \cite{lasota1985probabilistic}
A family of maps that are related to the cat maps are the perturbed cat maps,  \cite{boasman1995semiclassical, backer2003numerical} and their well-known form reads
\begin{eqnarray}
    &\begin{pmatrix}
        q' \\
        p'
    \end{pmatrix}
    =
    \begin{pmatrix}
        a & b \\
        c & d
    \end{pmatrix}
    \begin{pmatrix}
        q \\
        p
    \end{pmatrix}\nn \\
    &+
    \frac{\kappa}{2\pi}\cos(2\pi q)
    \begin{pmatrix}
        b \\
        d
    \end{pmatrix}
    {\rm mod\spa 1}.
\label{eq:genpertcatmap}
\end{eqnarray}
While $(a, b, c, d)$ can be arbitrary integers, in this work we will use a commonly studied example of $(a, b, c, d) = (2, 1, 3, 2)$.
Such a map is known to be fully chaotic for $\kappa \leq 0.33$.\cite{backer2003numerical}
Since $q$ and $p$ are the coordinates on a torus, it is useful to rewrite the map Eq.~(\ref{eq:gencatmap}) in terms of analytic variables on a torus $x$ and $z$, defined as
\begin{equation}
    z = e^{2 \pi i q} \;\; x = e^{2 \pi i p}.
\label{eq:compactvars}
\end{equation}
In these variables, the perturbed cat map of Eq.~(\ref{eq:genpertcatmap}) reads
\begin{eqnarray}
    &&x^\prime = x^2 z^3 \exp\left({\frac{i \kappa}{2}(z + z^{\mm})}\right) \nn \\
    &&z^\prime = x z^2 \exp\left({\frac{i \kappa}{2}(z + z^{\mm})}\right) 
\label{eq:classicalcatmapcompact}
\end{eqnarray}
An alternate way to specify the maps on a torus is in terms of a generating function $S(q', q)$, such that
\begin{equation}
    \hspace{-5mm}p = -\frac{\partial S(q', q)}{\partial q}\;\;\; p' = \frac{\partial S(q', q)}{\partial q'},
\label{eq:generatingfunctiondef}
\end{equation}
where $q$ and $q'$ are treated as independent variables. The generating function will be useful for the quantization of the map in the next section. For example, the generating function for the perturbed cat maps defined in Eq.~(\ref{eq:gencatmap}) reads
\begin{equation}
    S(q, q') = \frac{d {q'}^2}{2 b} - \frac{q q'}{b} + \frac{a q^2}{2 b} + \frac{\kappa}{4\pi^2}\sin(2\pi q).
\label{eq:perturbedcatgenerating}
\end{equation}

\subsection{Quantum Maps}
Since the perturbed cat maps are area-preserving, one can quantize the maps on a torus. \cite{berry1979quantum, hannay1980quantization, dematos1995quantization, balazs1989quantized, saraceno1990classical, degli1993quantization, backer2003numerical} While there are several approaches to quantization, we follow the one described in Ref.~[\onlinecite{backer2003numerical}].
We first define a Hilbert space that is compatible with the topology of a torus.
That is, the wavefunctions $\psi(q)$ and $\widetilde{\psi}(p)$ satisfy
\begin{equation}
    \psi(q + 1) = \psi(q),\;\;\; \widetilde{\psi}(p + 1) = \widetilde{\psi}(p),
\label{eq:torushilbertspace}
\end{equation}
resulting in the quantization of $p = j/N$ and $q = k/N$ for some $j, k \in \mathbb{N}$.\cite{backer2003numerical} 
Alternately, this constraint can be interpreted as the Planck's constant being allowed to only take values $\hbar = \frac{1}{2\pi N}$ for some $N \in \mathbb{N}$.
The semi-classical limit $\hbar \rightarrow 0$ thus corresponds to $N \rightarrow \infty$.
This $N$-dimensional Hilbert space is identical to that of a single particle moving on a one-dimensional periodic lattice with $N$ sites.
Since position and momentum operators do not commute, the phase space can be viewed as an $N \times N$ grid with each cell depicting the uncertainty of the position and momentum.
However, note that in contrast to physical particles moving on a periodic lattice, the maps we work with do not have any locality on the lattice. 
Alternately, the Hilbert space is that of the lowest Landau level in a quantum Hall system on a torus with $N$ flux quanta.\cite{fradkin2013field}

Since the Hilbert space is finite-dimensional, one can define a discrete position basis $\{\ket{q_j}\}$, where $q_j = j/N$.
In the position basis, the matrix elements of the unitary $U_N$ that describes the quantum map read \cite{backer2003numerical}
\begin{eqnarray}
    (U_N)_{{j'}, j} &\equiv& \bra{q_{j'}} U_N \ket{q_j} \nn \\
    &=& \frac{1}{\sqrt{N}}{\left\vert \frac{\partial^2 S(q', q)}{\partial q' \partial q}\right\vert^{1/2}_{q_{j'}, q_j}}\hspace{-5mm}\exp(2 \pi i N S(q_{j'}, q_j))
\label{eq:generalunitary}
\end{eqnarray}
where $S(q', q)$ is the generating function that satisfies the properties of Eq.~(\ref{eq:generatingfunctiondef}).
Using Eqs.~(\ref{eq:perturbedcatgenerating}) and (\ref{eq:generalunitary}) for the perturbed cat maps, $U_N$ reads
\begin{widetext}
\begin{equation}
    \hspace{-4mm}(U_N)_{j', j} = \frac{1}{\sqrt{Nb}} \exp\left(\frac{2 \pi i}{N b} \left(\frac{d {j'}^2}{2} - j j' + \frac{a j^2}{2}\right) + \frac{i \kappa N}{2 \pi}\sin\left(\frac{2 \pi j}{N}\right)\right)
\label{eq:unitarycatmap}
\end{equation}
\end{widetext}
While $U_N$ is not guaranteed to be unitary unless it satisfies a certain ``Egorov" property,\cite{backer2003numerical, frohlich2007semi, horvat2007egorov} we numerically find that when $b = 1$, $U_N$ is unitary for several valid values of $(a, c, d)$.
As mentioned before, we view the unitary of Eq.~(\ref{eq:unitarycatmap}) as a fundamental quantum system that has the classical limit of Eq.~(\ref{eq:genpertcatmap}) as $N \rightarrow \infty$. 
Chaos in the quantized maps is usually detected by the eigenvalue spacing statistics.\cite{bohigas1984characterization, backer2003numerical} That is, if $\lambda_n = e^{i \phi_n}$ are the eigenvalues of the unitary $U_N$, the distribution $P(s)$ of the nearest neighbor level spacings $s_n = (\phi_{n+1} - \phi_n)N/2\pi$ ($\phi_N \equiv \phi_0$) differs for a chaotic and a non-chaotic quantum map. 
For example, in the perturbed cat map for $\kappa \leq 0.33$, the distribution is expected to exhibit level repulsion, and is described by the Circular Orthogonal Ensemble (COE), which, in the $N \rightarrow \infty$ limit is the same as the Gaussian Orthogonal Ensemble (GOE). \cite{backer2003numerical}
While we numerically observe that the nature of the level statistics fluctuates significantly with $N$, we typically find that when $N$ is prime, the perturbed cat map exhibits GOE level statistics.  
\section{Heisenberg Picture and Equations of Motion}\label{sec:heisenbergpicture}
\subsection{Basis}
Quantum maps in the semi-classical limit are typically studied in a basis of coherent states (minimum uncertainty wavepackets) $\{\ket{q\spa p}\}$.\cite{kowalski1996coherent, kowalski2007coherent, fremling2014coherent}
In the $\hbar \rightarrow 0$ ($N \rightarrow \infty$) limit, the dynamics of these coherent states mimic the dynamics of individual points in phase space.
Several approaches to construct such sets of states for a particle on a finite-dimensional periodic lattice or for a quantum Hall system on a torus are known.\cite{bang2009wave, boon1978discrete, gonzalez1998coherent}
However, it is known to be impossible to obtain an $N$-dimensional orthogonal basis (for finite $N$) of wavefunctions on a torus that is local in phase space,\cite{zak1997balian} although orthogonalization of coherent states can be implemented in the continuum as well as the $N \rightarrow \infty$ limit. \cite{fremling2014coherent, rashba1997orthogonal}
This hinders an analytical exploration of the classical ($N \rightarrow \infty$) and quantum (finite $N$) dynamics in the same language in the Schr\"{o}dinger picture, although phase-space representations of quantum mechanics\cite{polkovnikov2010phase} such as the Wigner quasi-probability representations of wavefunctions (density matrices) have been employed. \cite{berry1979evolution, takahashi1985chaos, zurek1995quantum} 
However, as we show, operators in the Heisenberg picture have natural classical analogues that can be studied analytically.
Since the operator Hilbert space in this system is $N^2$-dimensional, a ``local" operator basis $\{\ket{q \spa p}\bra{q \spa p}\}$ constructed using the coherent states exists, and while $\{\ket{q\spa p}\}$ is an overcomplete basis of states, $\{\ket{q\spa p}\bra{q\spa p}\}$ is a complete basis of operators.
While such an operator basis is not orthogonal in general, we believe that an $N^2$-dimensional local orthogonal basis $\{P_{q,p}\}$ that resembles the coherent state operator basis $\{\ket{q \spa p}\bra{q \spa p}$ can be constructed. 
In the semi-classical limit, each operator basis element $P_{q_0,p_0}$ can be identified with $\delta(q - q_0, p - p_0)$ that corresponds to a $\delta$-function at a point $(q_0, p_0)$ in the phase space of the classical problem.
To obtain a basis that can be studied analytically, we use a compact ``position" operator $Z = \sum{e^{i 2 \pi q} P_{q,p}}$ and ``momentum" operator $X = \sum{e^{i 2 \pi p} P_{q,p}}$ to generate an $N^2$-dimensional ``Fourier" basis $\{X^m Z^n\}$ for any finite $N$. 
The semiclassical limit of each basis element $X^m Z^n$ is the classical Fourier ``basis element" $x^m z^n$, where $x$ and $z$ are defined in Eq.~(\ref{eq:compactvars}).
With this identification, we have obtained a natural analogue of quantum operators in the classical limit, complex-valued functions on phase space. 
While this correspondence has been known for long,\cite{polkovnikov2010phase, wong1998weyl} we study the equations of motion of quantum maps in such Fourier bases to analytically study classical and quantum dynamics. 
\subsection{Equations of Motion}
In an $N$-dimensional Hilbert space, the natural representation of the position and momentum operators $Z$ and $X$ are the $\mathbb{Z}_N$ clock operators that obey 
\begin{equation}
    X^N = Z^N = \mathds{1}\;\;\; XZ = \omega ZX, \;\;\; \omega \equiv e^{\frac{2\pi i}{N}}.
\label{eq:XZproperties}
\end{equation}
Using the expression for the unitary of the perturbed cat map in Eq.~(\ref{eq:unitarycatmap}), the Heisenberg equations of motion for $X$ and $Z$ is derived in App.~\ref{app:heisenbergeqnderivation}.
For the perturbed cat map with $(a, b, c, d) = (2, 1, 3, 2)$, they read (see Eqs.~(\ref{eq:Zpheis}) and (\ref{eq:Xpheis}))
\begin{eqnarray}
    &&X' = \omega^{\text{-} 3} X^2 Z^3 \exp\left(\frac{\kappa N}{4 \pi}(\omega - \omega^{\mm}) (\omega^{\mm} Z + \omega Z^{\mm})\right) \nn \\
    &&Z' = \omega^{\mm} X Z^2 \exp\left(\frac{\kappa N}{4 \pi}(\omega^{\frac{1}{2}} - \omega^{\text{-} \frac{1}{2}})(\omega^{\text{-} \frac{1}{2}} Z + \omega^{\frac{1}{2}}Z)\right), \nn \\
\label{eq:pertcatheisenbergeqs}
\end{eqnarray}
where $X'$ and $Z'$ are the time-evolved $X$ and $Z$ operators. 
In the limit of $N \rightarrow \infty$ ($\omega \rightarrow 1$), using the fact that $N (\omega^j - \omega^{\text{-} j}) \rightarrow 4 \pi i j$, Eq.~(\ref{eq:pertcatheisenbergeqs}) reduces to the evolution equation of $x$ and $z$'s in Eq.~(\ref{eq:classicalcatmapcompact}). 
Using the Heisenberg equations of motion, we derive the evolution of arbitrary operators written in the ``Fourier" basis.
We start with an operator $\mathcal{O}$, defined as 
\begin{equation}
    \mathcal{O} = \sum_{m,n}{Q_{m,n} X^m Z^n}.
\label{eq:quantumoperator}
\end{equation}
The operator evolution equation can be written as
\begin{equation}
    \mathcal{O}' = \sum_{m,n}{Q_{m,n} {X'}^m {Z'}^n}  \equiv \sum_{m,n}{Q'_{m,n} X^m Z^n},
\label{eq:quantumcoeffevoldefn}
\end{equation}
where $\mathcal{O}'$ is the time-evolved operator.
The relation of the time-evolved operator coefficients $\{Q'_{m,n}\}$ to the initial coefficients $\{Q_{m,n}\}$ is derived using Eq.~(\ref{eq:pertcatheisenbergeqs}).
As derived in App.~\ref{app:operatorevol} (see Eq.~(\ref{eq:pertcatmapquantcoeffapp})), the explicit evolution equation can be written as a matrix equation:
\begin{equation}
	\boldsymbol{Q'} = \boldsymbol{M}_Q \boldsymbol{Q}
\label{eq:quantummatrix}
\end{equation}
where $\boldsymbol{Q}$ ($\boldsymbol{Q'}$) is the vector of quantum coefficients $\{Q_{m,n}\}$ ($\{Q'_{m, n}\}$), and $\bM_Q$ is adjoint evolution (super-)operator with matrix elements (see Eq.~(\ref{eq:qmatrixelements})) 
\begin{eqnarray}
	&&\hspace{-4mm}\bra{m', n'} \bM_ Q \ket{m, n} = \omega^{-3m^2 - n^2 - 3mn}\sumal{s = 0}{N-1}{\left(\delta^{(N)}_{m', 2m + n} \delta^{(N)}_{n', 3m + 2n + s}\right.} \times  \nn \\
	&&\hspace{-5mm}\left.\sumal{p = -\infty}{\infty}{i^{s + pN} \mJ_{s + pN}\left(\frac{\kappa N}{\pi} \sin\left(\frac{\pi}{N}(2m + n)\right)\right) \omega^{\left(m + \frac{n}{2}\right)(pN - s)}}\right), \nn \\ 
\label{eq:adjointmatrixel}
\end{eqnarray}
where $\mJ_\nu(x)$ is the $\nu$-th Bessel function of the first kind and $\delta^{(N)}_{a,b} = 1$ if and only if $a = b\ \textrm{mod}\ N$, else 0. 
Since $\bM_Q$ is a quantum evolution operator, it is a unitary $N^2\times N^2$-dimensional matrix.
As discussed earlier, the classical analogue of Heisenberg evolution is the evolution of $L^\infty$ functions on phase space,  
\begin{equation}
    \mathcal{F} = \sum_{m,n}{C_{m,n} x^m z^n}.
\end{equation}
The evolution of the classical coefficients can then be computed using Eq.~(\ref{eq:classicalcatmapcompact}) 
\begin{equation}
    \mathcal{F}' = \sum_{m,n}{C_{m,n} {x'}^m {z'}^n} = \sum_{m,n}{C'_{m,n} x^m z^n},
\label{eq:classicalcoeffevoldefn}
\end{equation}
where $\mathcal{F}'$ is the time-evolved function. 
The set of time-evolved coefficients $\{C'_{m,n}\}$ is related to the initial coefficients $\{C_{m,n}\}$ using a matrix evolution equation similar to Eq.~(\ref{eq:quantummatrix}) (see Eq.~(\ref{eq:pertcatmapclasscoeffapp})),
\begin{equation}
	\boldsymbol{C'} = \boldsymbol{M}_C \boldsymbol{C},
\label{eq:classicalmatrix}
\end{equation}
where $\boldsymbol{C}$ ($\boldsymbol{C'}$) is the (infinite-dimensional) vector of classical coefficients $\{C_{m,n}\}$ ($\{C'_{m,n}\}$) and $\boldsymbol{M}_C$ is the evolution operator with matrix elements (see Eq.~(\ref{eq:cmatrixelements}))
\begin{eqnarray}
	\bra{m', n'} \bM_C \ket{m,n} = \sumal{s = -\infty}{\infty}{\left[i^s \mJ_{s}(\kappa(2m + n))\right.} \nn \\
	\left.\times \delta_{m', 2m + n} \delta_{n', 3m + 2n + s}\right].
\label{eq:koopmanmatrixel}
\end{eqnarray}
In Eq.~(\ref{eq:koopmanmatrixel}), one might recognize that $\bM_C$ is the classical Koopman operator written in the Fourier basis. \cite{koopman1931hamiltonian, lasota1985probabilistic}
Thus, as noted in previous works,\cite{koopman1931hamiltonian, peres2001hybrid} the quantum analogue of the Koopman operator $\bM_C$ is the adjoint evolution operator $\bM_Q$. 
The classical-quantum correspondence is established by studying the evolution of quantum coefficients in a $\mathbb{Z}_N \times \mathbb{Z}_N$ ``phase space" to the evolution of classical coefficients in a $\mathbb{Z} \times \mathbb{Z}$ space.
Moreover, since the total weights of the coefficients ($\sum_{m,n}{|Q_{m,n}|^2}$ and $\sum_{m,n}{|C_{m,n}|^2}$) are conserved (by virtue of unitary evolution), one can view this problem as the evolution of weights in the Fourier phase space. 
\section{Classical Evolution and Emergent Classicality}\label{sec:cqevolution}
In this section, we provide an operator interpretation of the \emph{classicality} of evolution and show that classicality can arise in quantum problems even away from the usual classical limit.   
In the classical map,  every point $(q, p)$ on phase space is mapped onto one other point $(q', p')$ on phase space.
In the classical limit of the quantum map, this property can be interpreted as the evolution of a basis element $P_{q,p} \sim \ket{q\ p}\bra{q\ p}$ into a different basis element $P_{q', p'} \sim \ket{q'\ p'}\bra{q'\ p'}$.
The existence of such a ``special" basis in which time-evolution is merely a shuffling of basis elements (without any phases) is interpreted as the \emph{classicality} of an evolution. 
That set of basis elements is in one-to-one correspondence with the ``phase space" of the system. For example, in the usual classical limit, the phase space is the infinite set $\{P_{q,p}\}$, or equivalently the set of all points $\{(q, p)\}$ on a torus.
Properties of classical evolution require the definition of a metric or a measure over this phase space.
For example, in classical maps, the set of basis elements is $\{P_{q,p}\}$ and the associated metric is the Euclidean distance on the usual phase space. 
Notions of ergodicity, classical chaos, mixing or exactness correspond to various behaviors of shuffling of these basis elements. 
For example, \emph{classical chaos} is the exponential separation of two nearby basis elements (with respect to the defined metric on phase space).
Similarly, a classical system is said to be \emph{ergodic} when the evolution of a single basis goes through all or almost all basis elements over a long time.
While such a ``special" basis does not exist for typical quantum systems, certain quantum systems exhibit ``emergent" classicality, as we now show for the quantized unperturbed Arnold cat map.  
We focus on the evolution of coefficients in the unperturbed cat map ($\kappa = 0$) of Eq.~(\ref{eq:gencatmap}) with $(a,b,c,d) = (2,1,3,2)$.
In this limit, the classical and quantum coefficient evolution matrix elements Eqs.~(\ref{eq:adjointmatrixel}) and (\ref{eq:koopmanmatrixel}) reduce to 
\begin{eqnarray}
    &&\bra{m', n'}\bM_Q \ket{m,n} = \omega^{-3 m^2 - n^2 - 3mn} \delta^{(N)}_{m', 2m + n} \delta^{(N)}_{n', 3m + 2n} \nn \\
    &&\bra{m', n'}\bM_C \ket{m,n} = \delta_{m', 2m + n} \delta_{n', 3m + 2n}.
\label{eq:kappa0mat}
\end{eqnarray}
In terms of coefficient evolution, Eq.~(\ref{eq:kappa0mat}) reduces to 
\begin{eqnarray}
    &&C'_{2m + n, 3m + 2n} = C_{m,n}, \nn \\
    &&Q'_{(2m + n, 3m + 2n)\ \textrm{mod}\ N} = \omega^{-3m^2 - n^2 - 3mn} Q_{m,n}.
\label{eq:unperturbed}
\end{eqnarray}
In Eq.~(\ref{eq:unperturbed}), each coefficient evolves into one other coefficient (or equivalently, each basis element evolves into one other basis element).
This is an example of classicality in the Fourier basis $\{x^m z^n\}$. 
The quantum operator coefficients behave similarly up to a phase that is picked up at each step of the evolution. 
However, since the phase is commensurate (i.e. of the form $e^{2\pi i p/q}$, where $p, q \in \mathbb{Z}$), the quantum evolution exhibits \emph{emergent classicality} on timescales of $\mathcal{O}(N)$ or lesser. 
As a consequence, notions of classical ergodicity and classical chaos can be applied to these systems.
Note that for generic finite-dimensional quantum systems, even though there always exists an eigenstate basis in which the operator evolution is merely a multiplication by a phase, the phases that the basis elements acquire are incommensurate, i.e. they are of the form $e^{2 \pi i r}$, where $r$ is irrational. Thus such cases do not qualify as emergent classicality.
Quantum systems that exhibit emergent classicality and have finitely-many degrees of freedom have a finite recurrence time, and thus the spectrum of the unitary is entirely composed of roots of unity. 
For example the quantized Arnold cat map with an $N$-dimensional Hilbert space,  the recurrence time is known to be $CN$,\cite{keating1991cat} where $C$ is an $\mathcal{O}(1)$ constant.
Indeed, such systems typically do not exhibit any of the usual signs of quantum chaos such as level repulsion or Wigner-Dyson statistics, as noted for the quantized cat map at $\kappa = 0$. \cite{hannay1980quantization, gamburd2003eigenvalue, keating1991cat}
Another set of well-known examples of quantum systems that exhibit emergent classicality are Clifford circuits, \cite{gottesman1998heisenberg, gutschow2010entanglement, gutschow2010time} where, in spite of absence of level repulsion, alternate notions of chaos and ergodicity apply in certain cases.\cite{von2017operator} 
It is not clear whether these emergent classical systems are \emph{quantum integrable}, since usual definitions of integrability in quantum mechanics hold only in systems with an infinite-dimensional Hilbert space (i.e. the thermodynamic limit in spin chains or the semi-classical limit for quantum maps), or for a continuous family of systems with a finite-dimensional Hilbert space.\cite{yuzbashyan2013quantum, yuzbashyan2016rotationally, scaramazza2016integrable, gritsev2017integrable} 
To summarize, the perturbed cat map exhibits classicality in three different limits.
First, in $N \rightarrow \infty$ limit when $\kappa \neq 0$, where the $\{P_{q,p}\}$ basis is the only basis in which the operator evolution is classical. 
Second, when $\kappa = 0$ for finite $N$, where the operator evolution is classical only in the $\{X^m Z^n\}$ basis.
Third, in the $N \rightarrow \infty$ limit, when $\kappa = 0$, where the operator evolution is classical in both the $\{P_{q, p}\}$ and $\{X^m Z^n\}$ bases.
Before moving on to quantum systems, two comments on the behavior of classically chaotic maps in the Fourier ($\{x^m z^n\}$) basis are in order, which we illustrate these using the evolution of classical coefficients in the $\kappa = 0$ limit.
Following Eq.~(\ref{eq:unperturbed}), an initial set of coefficients $\{C^{(0)}_{m,n}\}$ evolve into coefficients $\{C^{(t)}_{m,n}\}$ that do not vanish for larger $m$ and $n$, and at time $t \gg 1$ the initial and final coefficients are related by
\begin{equation}
    C^{(0)}_{m,n} \approx C^{(t)}_{e^{\lambda t}\left( \left[m/2 + n/(2\sqrt{3})\right],\left[(\sqrt{3}/2) m + n/2\right]\right)},
\label{eq:comcoefficient}
\end{equation}
where $\lambda = \log{(2 + \sqrt{3})}$, the classical Lyapunov exponent of the unperturbed cat map and $[\;]$ denotes the integer part. 
Firstly, according to Eq.~(\ref{eq:comcoefficient}), any initial smooth function on phase space thus evolves into a ``rough" function on phase space at late times.
The exponential growth of ``roughness" is \emph{classical chaos}, and its rate is characterized by the Lyapunov exponent $\lambda$. 
Secondly, the statement of classical ergodicity says that in an ergodic map, the only eigenfunction of the evolution operator $\bM_C$ (Koopman operator) is the uniform (constant) function.\cite{gaspard2005chaos}
In the Fourier basis, the time-evolution of coefficients in an ergodic classical map satisfy
\begin{equation}
    C^{(0)}_{0,0} = C^{(t)}_{0,0},
\label{eq:ergodicity}
\end{equation}
which is consistent with Eq.~(\ref{eq:comcoefficient}).
\section{Regimes of Quantum Evolution}\label{sec:regimes}
\begin{figure*}[ht]
    \begin{tabular}{cc}
        \includegraphics[scale = 0.5]{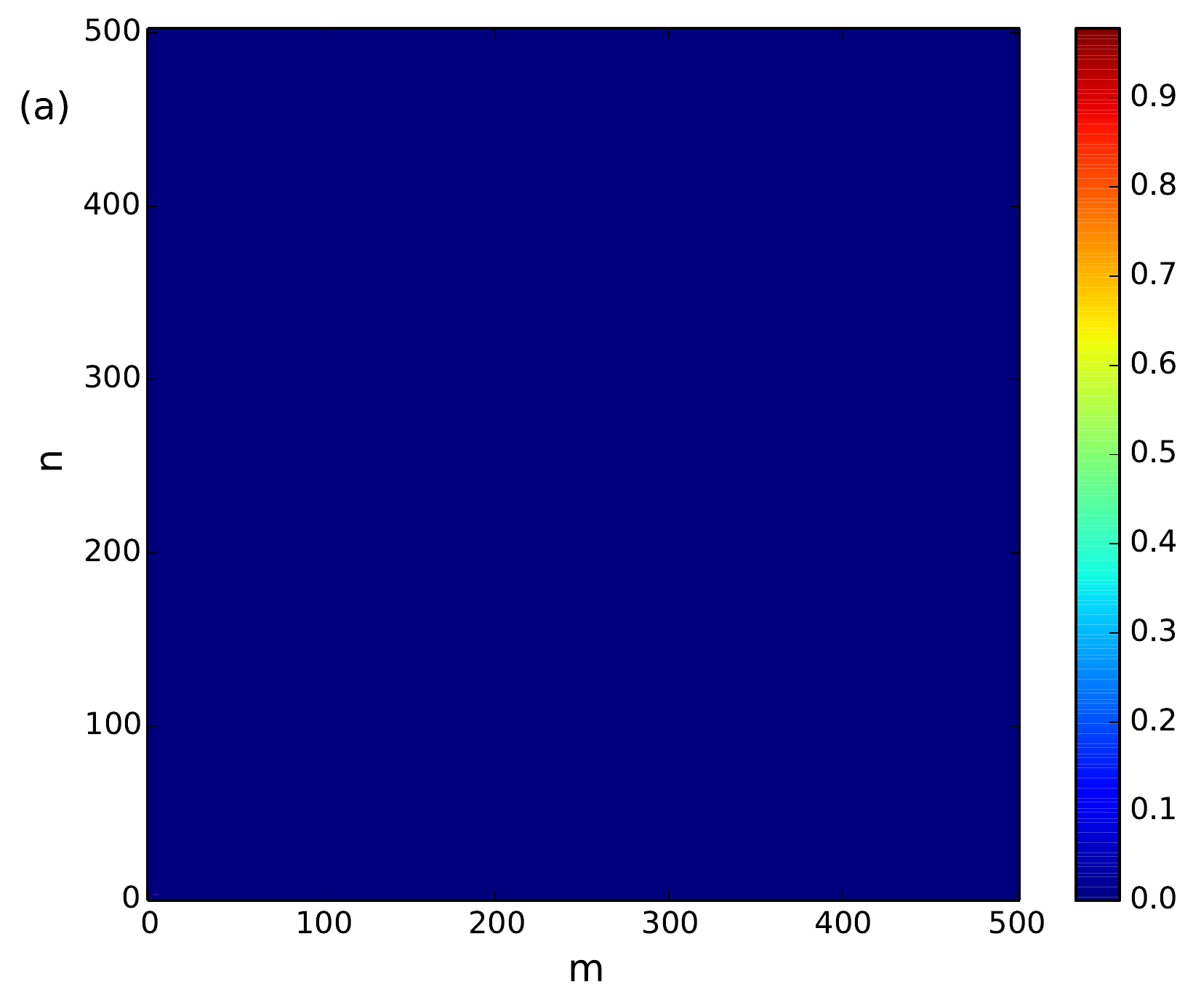}&\includegraphics[scale = 0.5]{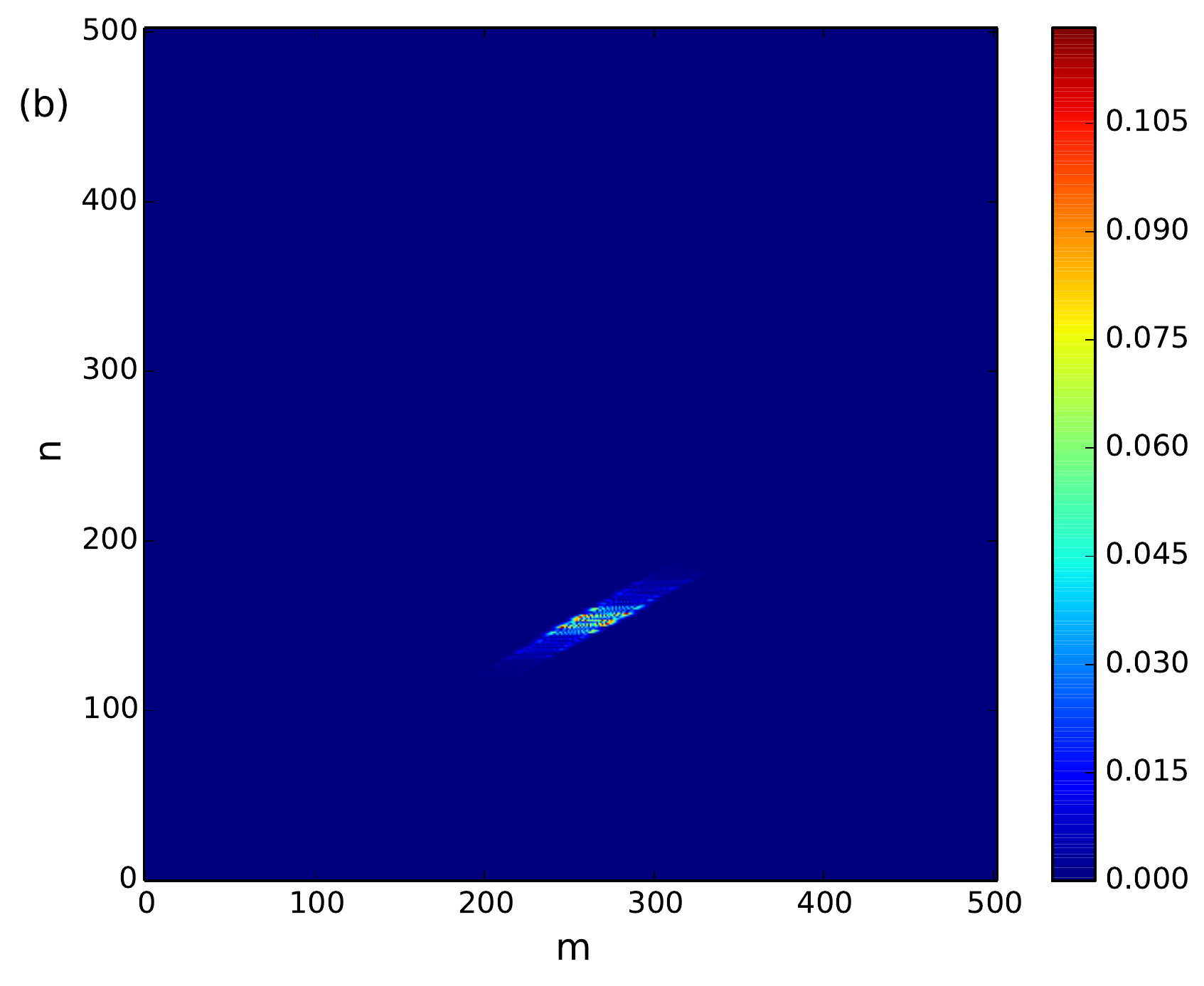}\\
        \includegraphics[scale = 0.5]{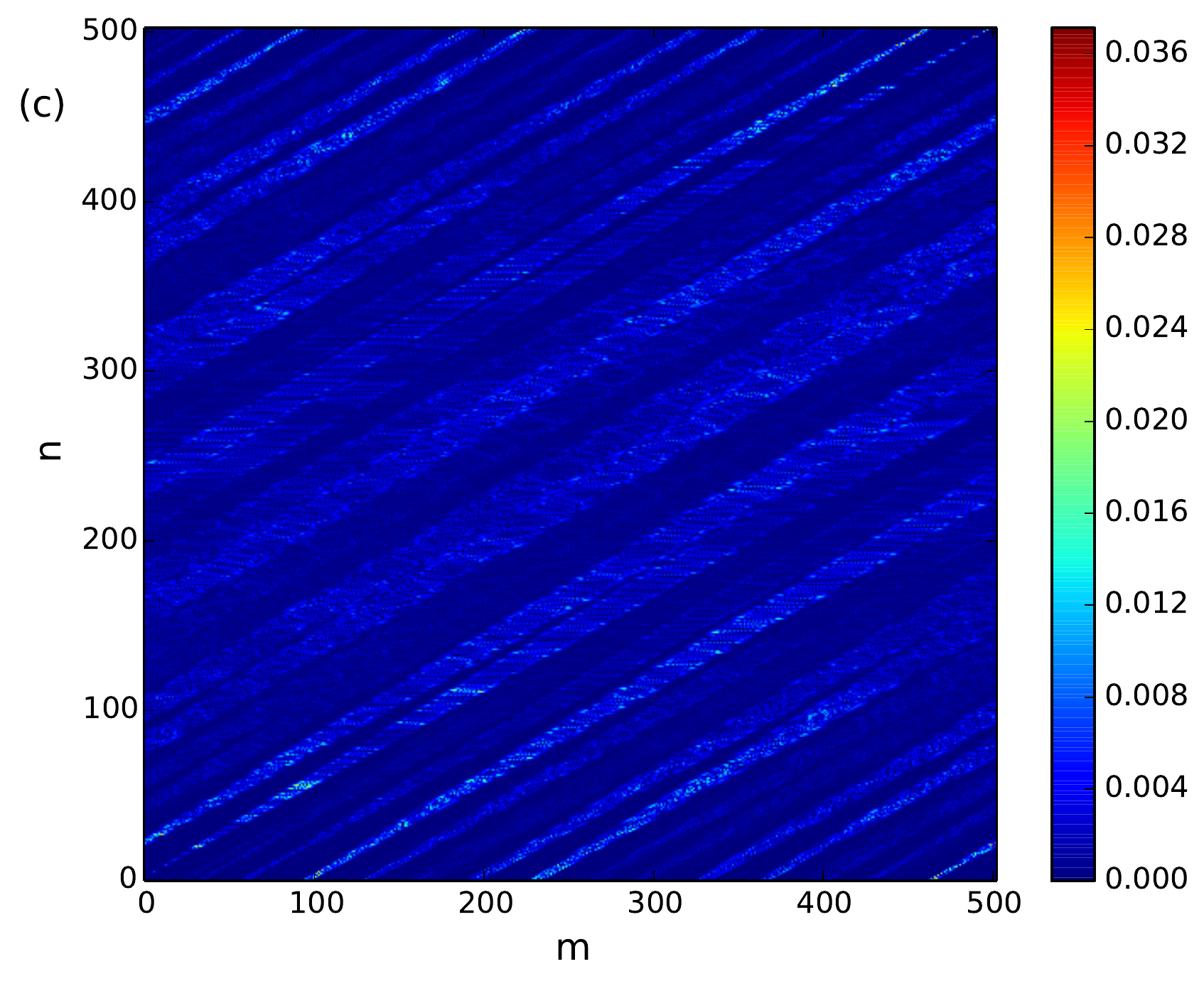}&\includegraphics[scale = 0.5]{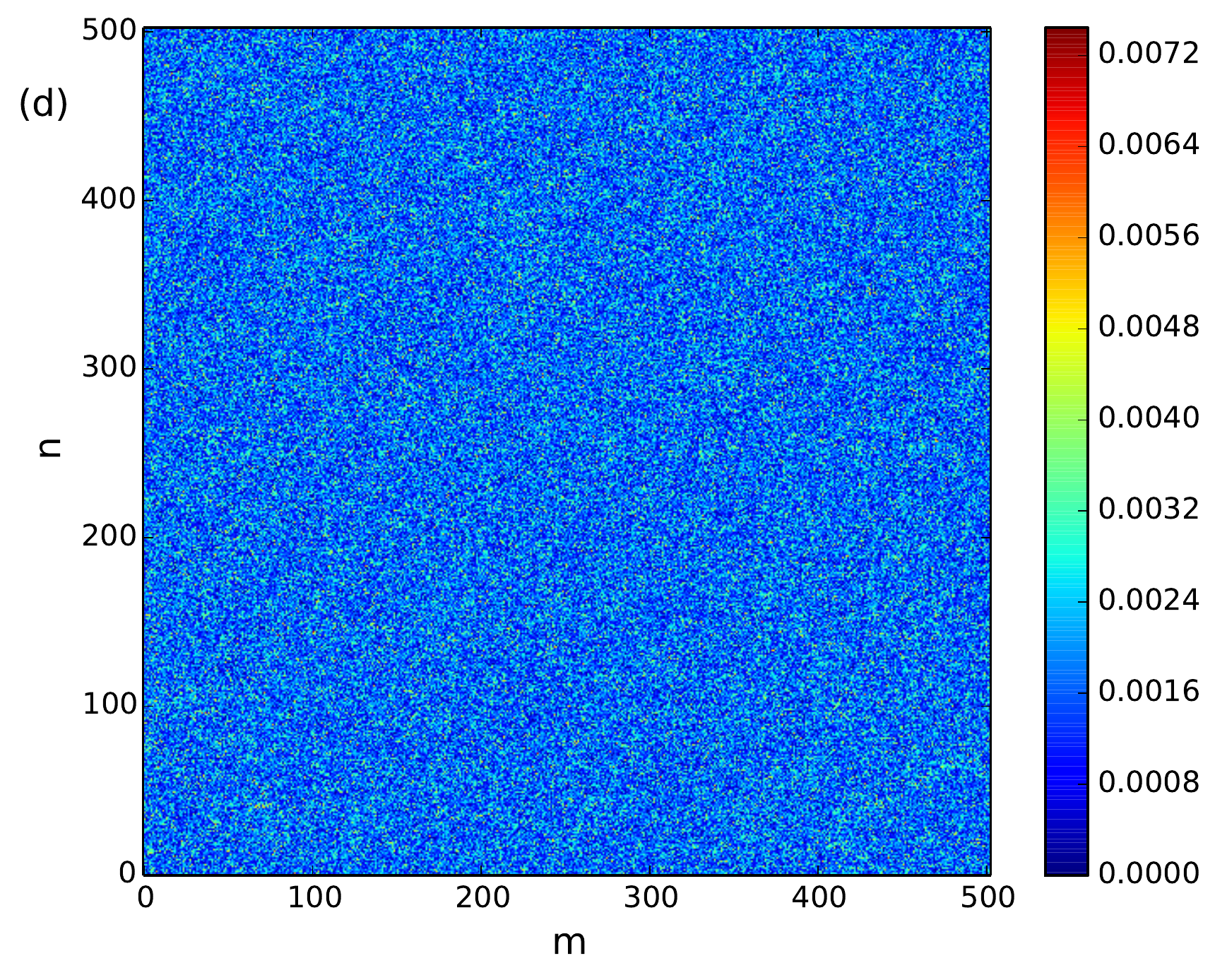}\\
    \end{tabular}
\caption{(Color online) Operator coefficient evolution of an initial operator $XZ$ for $\kappa = 0.1$ and $N = 503$ at various times. Note that the Fourier phase space has periodic boundary conditions. (a) $t = 0$: Initial operator (b) $t = 4$, Early times: For $t < t_E$ the operator coefficients are localized in the Fourier phase space.  (c) $t = 8$, Intermediate times: For $t_E < t < t_{\textrm{scr}}$ the localized set of operator coefficients spread enough to wrap around the Fourier phase space. (d) $t = 14$, Late times: For $t > t_{\textrm{scr}}$, the coefficients are essentially uniformly spread throughout the Fourier phase space. }
\label{fig:phasespace}
\end{figure*}
In a chaotic quantum system, we do not expect any special basis in which classicality emerges. 
That is, in general the operator coefficients spread in any basis.
We illustrate the nature of the evolution of coefficients in a chaotic quantum map.
Useful quantities to study the time-evolution of the coefficients are the classical and quantum entropies $S_C(t)$ and $S_Q(t)$ defined as
\begin{eqnarray}
    S_C(t) &\equiv& -\sum_{m,n}{|C^{(t)}_{m,n}|^2 \log |C^{(t)}_{m,n}|^2} \nn \\
    S_Q(t) &\equiv& -\sum_{m,n}{|Q^{(t)}_{m,n}|^2 \log |Q^{(t)}_{m,n}|^2},
\label{eq:entropies}
\end{eqnarray}
where $\{C^{(t)}_{m,n}\}$ and $\{Q^{(t)}_{m,n}\}$ are the sets of classical and quantum coefficients at time $t$. The coefficients are normalized such that
\begin{equation}
    \sum_{m,n}{|C^{(t)}_{m,n}|^2} = 1,\;\;\; \sum_{m,n}{|Q^{(t)}_{m,n}|^2} = 1.
\end{equation}
To establish a classical-quantum correspondence, we choose the initial classical functions and quantum operators such that the set of coefficients $\{C^{(0)}_{m,n}\}$ and $\{Q^{(0)}_{m,n}\}$ are the same and they satisfy the property
\begin{equation}
    Q^{(0)}_{m,n} = C^{(0)}_{m,n} = 0\;\; \textrm{if}\;\; m > m_0\;\; \textrm{or}\;\; n > n_0,\;\;\; m_0, n_0 \ll N.
\label{eq:initialcoeff}
\end{equation}
The property of Eq.~(\ref{eq:initialcoeff}) ensures the smoothness of the initial function on the classical phase space and, as we will see, enables the estimation of an Ehrenfest time for simple operators.
As discussed in the previous section, in the $\kappa = 0$ problem, every coefficient is mapped on to one other coefficient. 
Thus the initial set of coefficients do not spread in Fourier space.
Consequently, the classical and quantum entropies defined in Eq.~(\ref{eq:entropies}) are conserved ($S_C(t) = S_C(0), \;\; S_Q(t) = S_Q(0)\;\; \forall t$). 
However, when $\kappa \neq 0$, the quantum coefficients (and entropy) behave differently, and their evolution can be classified into three distinct regimes. 
Note that in this section we will only be concerned about the scaling of the timescales with $N$ and $\kappa$, not quantitatively accurate estimates. 
\begin{figure}[tb]
\includegraphics[scale=0.47]{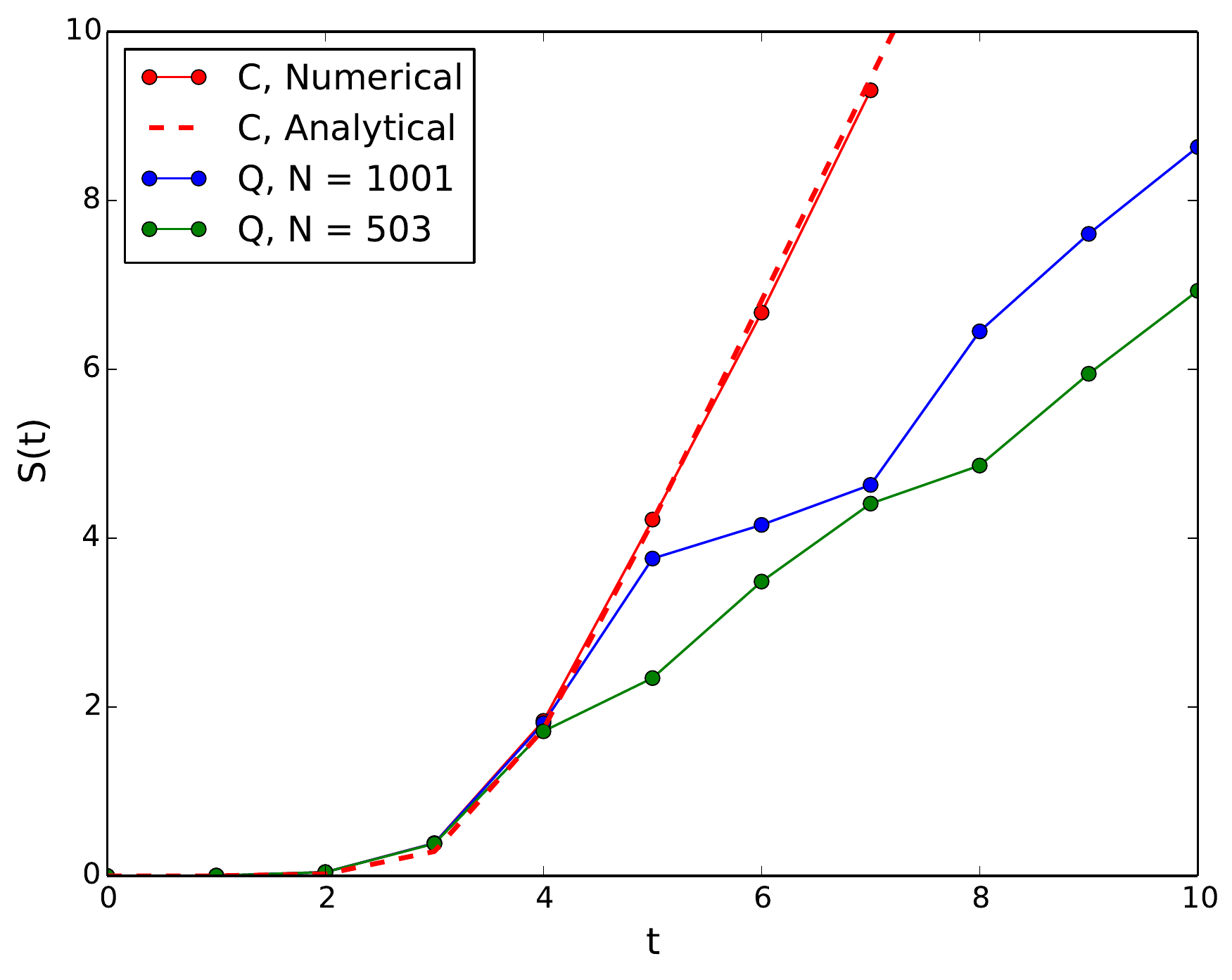}
\caption{(Color online) The classical (C) and quantum (Q) coefficient entropies of the early-time evolution of an initial operator $X Z$ in the perturbed cat map with $\kappa = 0.01$. The dashed line shows $S_C(t)$ of Eq.~(\ref{eq:earlyestimate}) with $c_0 \approx 5.88$ and $A_C(0) = 1$. Note that the classical and quantum entropies deviate after an Ehrenfest time that can be estimated (using Eq.~(\ref{eq:ehrenfesttime})) to be $t_E \approx 5.24$ for $N = 1001$.}
\label{fig:earlytime}
\end{figure}
\subsection{Early Times}\label{sec:ehrenfesttime}
At early times in the perturbed cat map for small $\kappa$, we expect a qualitative behavior similar to the unperturbed cat map.
For very small $\kappa$, each Fourier coefficient follows the mapping of Eq.~(\ref{eq:comcoefficient}), along with a small ``spreading" since for any $\kappa > 0$, i.e. a single basis element evolves into a superposition of several basis elements.
This is evident by writing down the matrix elements of Eq.~(\ref{eq:koopmanmatrixel}) in a more suggestive form 
\begin{equation}
	\bra{2m + n, 3m + 2n + s} \bM_C \ket{m, n} = {i^s \mJ_{s}\left(\kappa\left(2m + n\right)\right)}.
\label{eq:clmatrixapprox}
\end{equation}
As shown in Eq.~(\ref{eq:deb}) in App.~\ref{app:spreading}, $|\mJ_\nu(x)|$ can be considered to vanish when $|\nu| > |x|$ and $|x| \gg 1$. 
The typical spreading $\xi$ in the $n$-direction of a coefficient $(m,n)$ in Eq.~(\ref{eq:clmatrixapprox}) is thus 
\begin{equation}
    \xi \approx \kappa (2m + n).
\label{eq:spreadingestimateclassical}
\end{equation}
If the set of classical coefficients has an ``area" (number of non-zero coefficients) $A_C$, the approximate width in the $m$-direction is $\sqrt{A_C}$, and hence the change in the area at every step is given by 
\begin{eqnarray}
    \Delta A_C(t) &\approx& 2 \xi \sqrt{A_C(t-1)} = 2 \kappa(2m + n) \sqrt{A_C(t-1)} \nn \\
    &\approx& 2 c_0 \kappa e^{\lambda t} \sqrt{A_C (t-1)},\nn \\
\label{eq:classicalareachange}
\end{eqnarray}
where $c_0$ is a constant that depends on the initial choice of coefficients. In Eq.~(\ref{eq:classicalareachange}) we have used the fact that $m$ and $n$ for the set of non-vanishing coefficients grows exponentially with the Lyapunov exponent $\lambda$ (see Eq.~(\ref{eq:comcoefficient})).
Consequently, we can estimate the area and the entropy of the coefficients to be
\begin{eqnarray}
    A_C(t) &\approx& A_C(0) + \frac{c_0^2 \kappa^2}{\lambda^2} e^{2 \lambda t}\nn \\
    \implies S_C(t) &\approx& \log A_C(t) \approx \log\left(A_C(0) + \frac{c_0^2 \kappa^2}{\lambda^2} e^{2 \lambda t}\right)\nn \\
    &\sim& 2 \lambda t + \textrm{constant} 
\label{eq:earlyestimate}
\end{eqnarray}
We compare the classical and quantum evolutions by choosing a classical function on phase space and a quantum operator whose coefficients $\{C^{(0)}_{m,n}\}$ and $\{Q^{(0)}_{m,n}\}$ respectively are the same and they satisfy the property of Eq.~(\ref{eq:initialcoeff}).
For small $m_0, n_0$ and large $N$, the quantum coefficients mimic the classical coefficients since Eq.~(\ref{eq:adjointmatrixel}) reduces to Eq.~(\ref{eq:koopmanmatrixel}) in this limit.
Since Eqs.~(\ref{eq:adjointmatrixel}) and (\ref{eq:koopmanmatrixel}) differ only when $m, n \sim \mathcal{O}(N)$, the evolution of the quantum and classical coefficients differ only when the magnitudes of coefficients $Q^{(t)}_{m,n}$ and $C^{(t)}_{m,n}$ are significant for $m,n \sim \mathcal{O}(N)$. 
For a small $\kappa$, since all the coefficients evolve roughly according to Eq.~(\ref{eq:comcoefficient}) (up to small spreading), this timescale (Ehrenfest time $t_E$) can be estimated to be 
\begin{equation}
 e^{\lambda t_E}\textrm{max}\left(m_0, n_0\right) \sim N \implies  t_E \sim \frac{1}{\lambda}\log N,
\label{eq:ehrenfesttime}
\end{equation}
where we have assumed $\textrm{max}\left(m_0, n_0\right) \ll N$.
This behavior of the classical and quantum entropies is shown in Fig.~\ref{fig:earlytime}.
The area and the entropy of the coefficients at the Ehrenfest time are
\begin{eqnarray}
    &&A_Q(t_E) \approx A_C(t_E) \approx \frac{c_0^2 \kappa^2 N^2}{\lambda^2},\nn \\
    &&S_Q(t_E) \approx S_C(t_E) \approx 2 \log\left(\frac{c_0 \kappa N}{\lambda}\right).
\label{eq:ehrenfestentropy}
\end{eqnarray}
Since the quantum and classical coefficients are similar for $t < t_E$, classicality (as discussed in the previous section) approximately holds for the quantum problem in the $P_{q, p}$ basis. That is, a given basis element $P_{q, p}$ approximately evolves into $P_{q', p'}$.  
\subsection{Intermediate Times}\label{sec:intermediatetimes}
\begin{figure}[tb]
\includegraphics[scale=0.47]{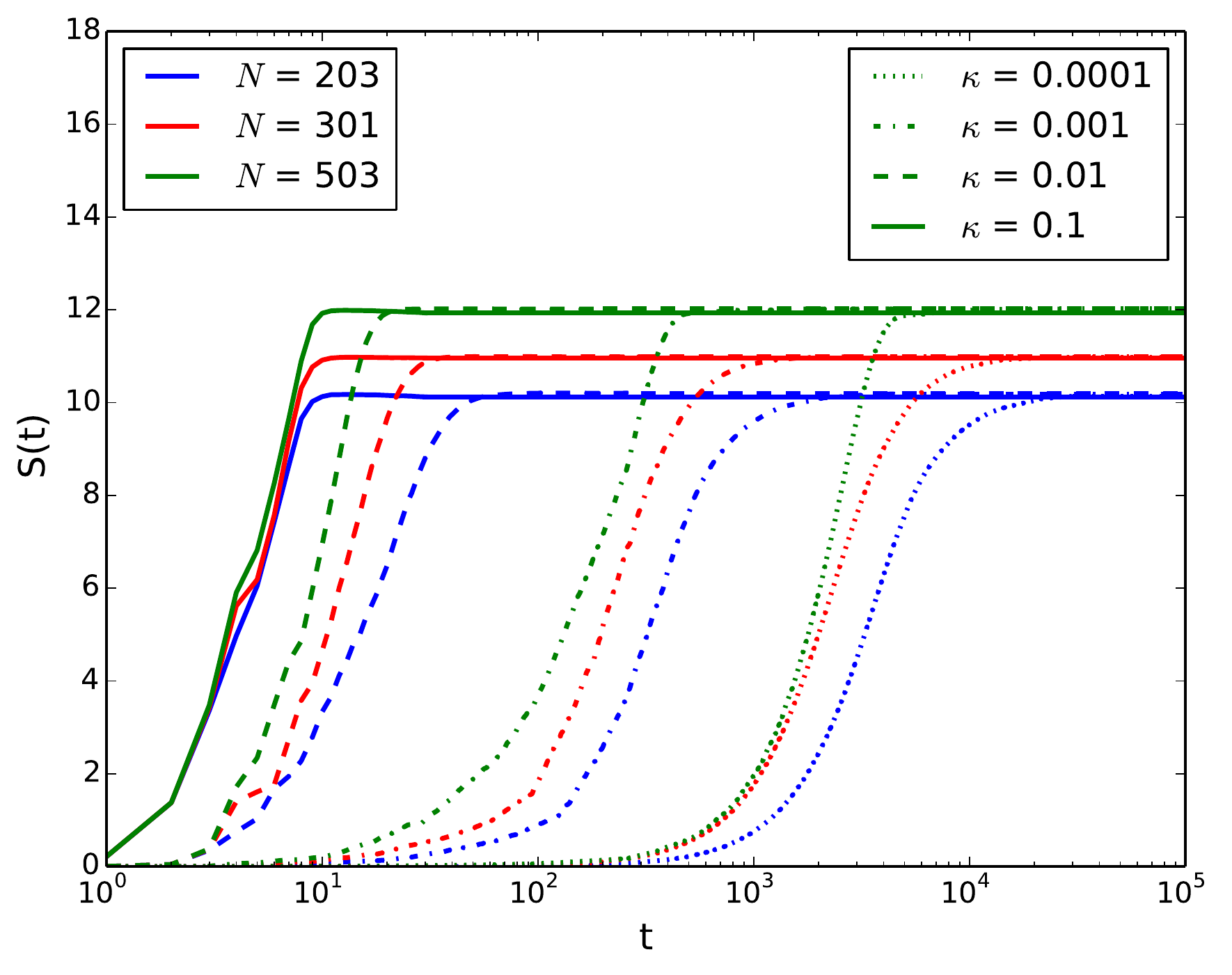}
\caption{(Color online) The quantum coefficient entropies of the evolution of an initial operator $X Z$ at intermediate and late times. Note that the entropy saturates at an operator scrambling time that obeys Eq.~(\ref{eq:scramblingtime}). }
\label{fig:intermediatetime}
\end{figure}
Once the evolution of quantum coefficients deviates from the classical evolution, there is a timescale until which the quantum operator undergoes Hamming spreading over the Fourier basis until the operator coefficients look random in this basis.
This is the time at which the growth of the quantum entropy $S_Q(t)$ saturates, for example as seen in Fig.~\ref{fig:intermediatetime}.
We call this the \emph{operator scrambling time}, the time at which the operator has a roughly uniform weight on each of the Fourier basis elements. 
To estimate this time-scale, we note that for $\kappa \ll 1$, Eq.~(\ref{eq:adjointmatrixel}) can be approximated to be 
\begin{eqnarray}
	&&\bra{2m + n, 3m + 2n + s} \bM_ Q \ket{m, n} \nn \\
	&&\approx i^{s} \mJ_{s}\left(\frac{\kappa N}{\pi} \sin\left(\frac{\pi}{N}(2m + n)\right)\right) \omega^{-s\left(m + \frac{n}{2}\right)}.
\label{eq:adjointmatrixel2}
\end{eqnarray}
Similar to the early-time case, the we assume a typical spreading $\xi$ in the $n$-direction.
Since the approximation for the Bessel function $\mJ_\nu(x)$ depends on whether $|x| \gg 1$ or $|x| \ll 1$ (see App.~\ref{app:spreading})) the typical spreading $s$ in Eq.~(\ref{eq:adjointmatrixel2}) can be estimated to be (see Eqs.~(\ref{eq:nu0smallx}) and (\ref{eq:nu0largex}))
\begin{equation}
    \xi \sim \twopartdef{\frac{1}{\log\left(\frac{1}{\kappa N}\right)}}{\kappa N \ll 1}{\kappa N}{\kappa N \gg 1}. 
\label{eq:spreadingvals}
\end{equation}
Similar to Eq.~(\ref{eq:classicalareachange}), the area of the coefficients obey 
\begin{equation}
    \Delta A_Q(t) \approx 2\xi \sqrt{A_Q(t-1)}.
\label{eq:quantumareachange}
\end{equation}
The area $A_Q(t)$ quantum coefficients thus grows quadratically,
\begin{eqnarray}
    &\sqrt{A_Q(t)} - \sqrt{A_Q(t_E)} \approx \xi (t - t_E)\nn \\
    &\implies A_Q(t) \approx \left(\frac{c_0 \kappa N}{\lambda} - \frac{\xi \log N}{\lambda} + \xi t\right)^2,
\end{eqnarray}
and the entropy $S_Q(t)$ grows logarithmically.
\begin{equation}
    S_Q(t) \approx \log A_Q(t) \approx 2 \log \left(\frac{c_0 \kappa N}{\lambda} - \frac{\xi \log N}{\lambda} + \xi t\right)
\end{equation}
This behavior continues until the area of the coefficients is $\mathcal{O}(N^2)$. Thus, the operator scrambling time can be estimated to be 
\begin{equation}
    t_{\textrm{scr}} \sim t_E + \frac{C N}{\xi}, 
\label{eq:scramblingtime}
\end{equation}
where $C$ is a constant and $t_E$ is the Ehrenfest time of Eq.~(\ref{eq:ehrenfesttime}). 
Substituting Eqs.~(\ref{eq:spreadingvals}) and (\ref{eq:ehrenfesttime}) in Eq.~(\ref{eq:scramblingtime}), we obtain 
\begin{equation}
    t_{\textrm{scr}} \sim \frac{\log N}{\lambda} + \twopartdef{N \log\left(\frac{1}{\kappa N}\right)}{\kappa N \ll 1}{\frac{1}{\kappa}}{\kappa N \gg 1}.
\label{eq:scrambtime}
\end{equation}
Thus, we expect a crossover between the two behaviors when $\kappa N \sim \mathcal{O}(1)$. 
The $\kappa N \ll 1$ regime in Eq.~(\ref{eq:scrambtime}) resembles the ``emergent classical" behavior of the $\kappa = 0$ limit of the perturbed cat map discussed in Sec.~\ref{sec:cqevolution} and is not representative of generic chaotic quantum systems. 
Note that a slightly different form of the operator scrambling time was conjectured in Ref.~[\onlinecite{chen2018operator}].
\subsection{Late Times}\label{sec:latetime}
After an operator scrambling time, a small set of initial operator coefficients evolves into one that has an equal weight on all of the basis elements of the semi-classical basis. 
That is, the quantum entropy $S_Q(t)$ nearly saturates the bound
\begin{equation}
    S_Q \leq 2 \log N.
\label{eq:quantumentropybound}
\end{equation}
This is a characteristic feature of quantum chaos, and this entropy saturation holds for most initial operators after the operator scrambling time.  
In particular, one can choose any initial operator $P_{pq} \sim \ket{q\ p}\bra{q\ p}$ (for any $q$ and $p$) and expect the operator coefficients of $U^\dagger P_{pq} U$ to be uniformly spread in operator space after an operator scrambling time. 
That is, 
\begin{eqnarray}
    \textrm{Tr}\left(\left(U^\dagger\right)^t \ket{q\ p}\bra{q\ p} U^t X \right) &\sim& \textrm{Tr}\left(\left(U^\dagger\right)^t \ket{q'\ p'}\bra{q'\ p'} U^t X \right) \nn \\
    \implies \bra{q\ p} U^t X \left(U^\dagger\right)^t \ket{q\ p} &\sim& \bra{q'\ p'} U^t X \left(U^\dagger\right)^t \ket{q'\ p'},
\label{eq:statescramblingcondition}
\end{eqnarray}
which shows that expectation values of $X$ (or any basis element) equilibriate to the same value at late times irrespective of the initial a coherent state $\ket{q\ p}$. 
Thus the operator scrambling time is the same as the \emph{state scrambling time}, the time at which any initial coherent state wavepacket centered at $(q, p)$ on phase space uniformly spreads throughout the system.
It is important to reconcile certain aspects of the late-time classical evolution discussed in Sec.~\ref{sec:cqevolution} and late-time quantum evolution.
The classical and quantum late-time behaviors are both governed by the eigenstates of $\bM_C$ and $\bM_Q$ that have an eigenvalue of unit magnitude. The outcomes differ in these cases. 
$\bM_Q$ has $N^2$ eigenstates with eigenvalue of magnitude 1. If $\{\ket{\phi_m}\}$ are eigenstates of the quantum unitary $U$ with eigenvalues $\{\exp(i \phi_m)\}$, $\{\ket{\phi_m}\bra{\phi_n}\}$ are the eigenstates of $\bM_Q$ with eigenvalues $\{\exp(i (\phi_m - \phi_n))\}$.
As discussed in Sec.~\ref{sec:cqevolution}, $\bM_C$ for classically chaotic systems has a single eigenfunction, which is the constant function on phase space. 
Thus, most ($N^2 -1$) quantum eigenstates do not have a clear meaning in the $N \rightarrow \infty$ limit and it has to be the case that all except one of the quantum eigenstates map onto singular functions on the classical phase space.
\section{Basis-Independence and the Spectral Form Factor}\label{sec:spectrum}
\begin{figure}[tb]
\includegraphics[scale=0.47]{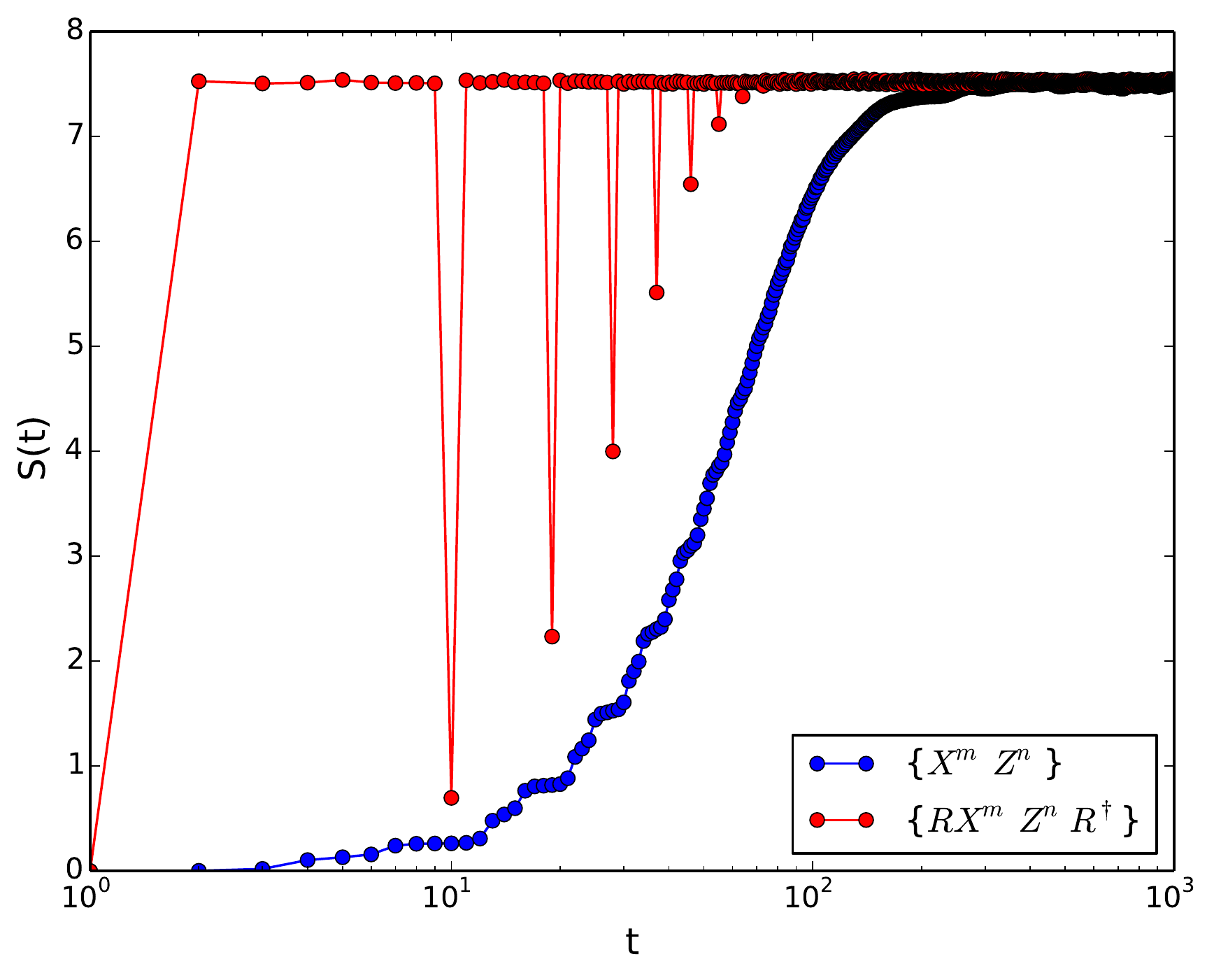}
\caption{(Color online) The quantum coefficient entropies of the evolution of an initial operator with a low entropy in the perturbed cat map with $\kappa = 0.01$ in the bases $\{X^m Z^n\}$ and $\{R X^m Z^n R^\dagger\}$, where $R$ is a randomly chosen unitary. We find that for several choices of $R$, the operator scrambling time is of the order of $t_{\textrm{scr}}$ computed analytically for the $\{X^m Z^n\}$ basis.}
\label{fig:randombasis}
\end{figure}
One might worry that the study of operator evolution with a different choice of basis  in Sec.~\ref{sec:regimes} (instead of $\{X^m Z^n\}$) might lead to a different behavior of the operator coefficients. 
In a fully quantum chaotic system, we find that for any generic choice of basis $\{R X^m Z^n R^\dagger\}$, where $R$ is a random unitary, the entropy of the operator coefficients $S_Q(t)$ does not equilibriate until a timescale of the operator scrambling time of Eq.~(\ref{eq:scramblingtime}), as shown in Fig.~\ref{fig:randombasis}.
However, the growth of $S_Q(t)$ is not monotonic in the $\{R X^m Z^n R^\dagger\}$ basis, and even though $S_Q(t)$ reaches its maximum value at early times, its shows recurrences to low values at early times. 
We thus believe that quantum chaos is characterized by the late-time saturation of entropy $S_Q(t)$ for most initial operators and most choices of operator bases. 
An important caveat for finite-dimensional quantum systems is that one can always choose the operator basis $\{\ket{\phi_m}\bra{\phi_n}\}$ formed by the eigenstates of the time-evolution unitary $U$ in which operator coefficients do not spread irrespective of whether the system is quantum chaotic or integrable.
The existence of this special basis is related to the difficulty of defining integrability and chaos in finite-dimensional systems.\cite{yuzbashyan2013quantum, yuzbashyan2016rotationally, scaramazza2016integrable}
Of course, in order to observe a classical-quantum correspondence and an Ehrenfest time in the evolution of operator coefficients, the choice of an operator basis is more restrictive. 
A quantum map is formally said to have a classical limit if it satisfies the Egorov condition, a strong version of which loosely states that in the semi-classical ($N \rightarrow \infty$) limit of the quantum map, the time-evolution of smooth functions on phase space and quantum operators commute (see Eq.~(13) of Ref.~[\onlinecite{backer2003numerical}]). 
Due to the restriction of the Egorov condition to the behavior of smooth functions on phase space, in order to establish a classical-quantum correspondence, it is natural to use a quantum basis that limits to a basis of smooth functions on phase space as a classical basis. 
Other choices of quantum bases do not have a clear meaning in the classical limit, at least from the perspective of the Egorov condition.
Thus, by imposing the condition that the quantum basis maps on to a basis of smooth functions in the classical limit, we rule out most choices of bases, for example the operator basis formed by eigenstates of the unitary (as discussed in Sec.~\ref{sec:latetime} not all of them can map on to smooth functions in the classical limit).
Clearly the quantum basis $\{X^m Z^n\}$ and the classical Fourier basis $\{x^m z^n\}$ itself satisfy the required properties. 
We believe the behavior of the coefficients and entropy should remain qualitatively the same with any other choice basis that satisfies the required properties, although it is not clear how to construct an analytical example that is different from $\{X^m Z^n\}$.
However, we note that weaker versions of the Egorov condition (see Ref.~[\onlinecite{bievre1998egorov}] for an example) could lead to alternate sensible choices of bases in both the quantum and classical limits, an interesting avenue for future work.
\begin{figure}[tb]
\includegraphics[scale=0.47]{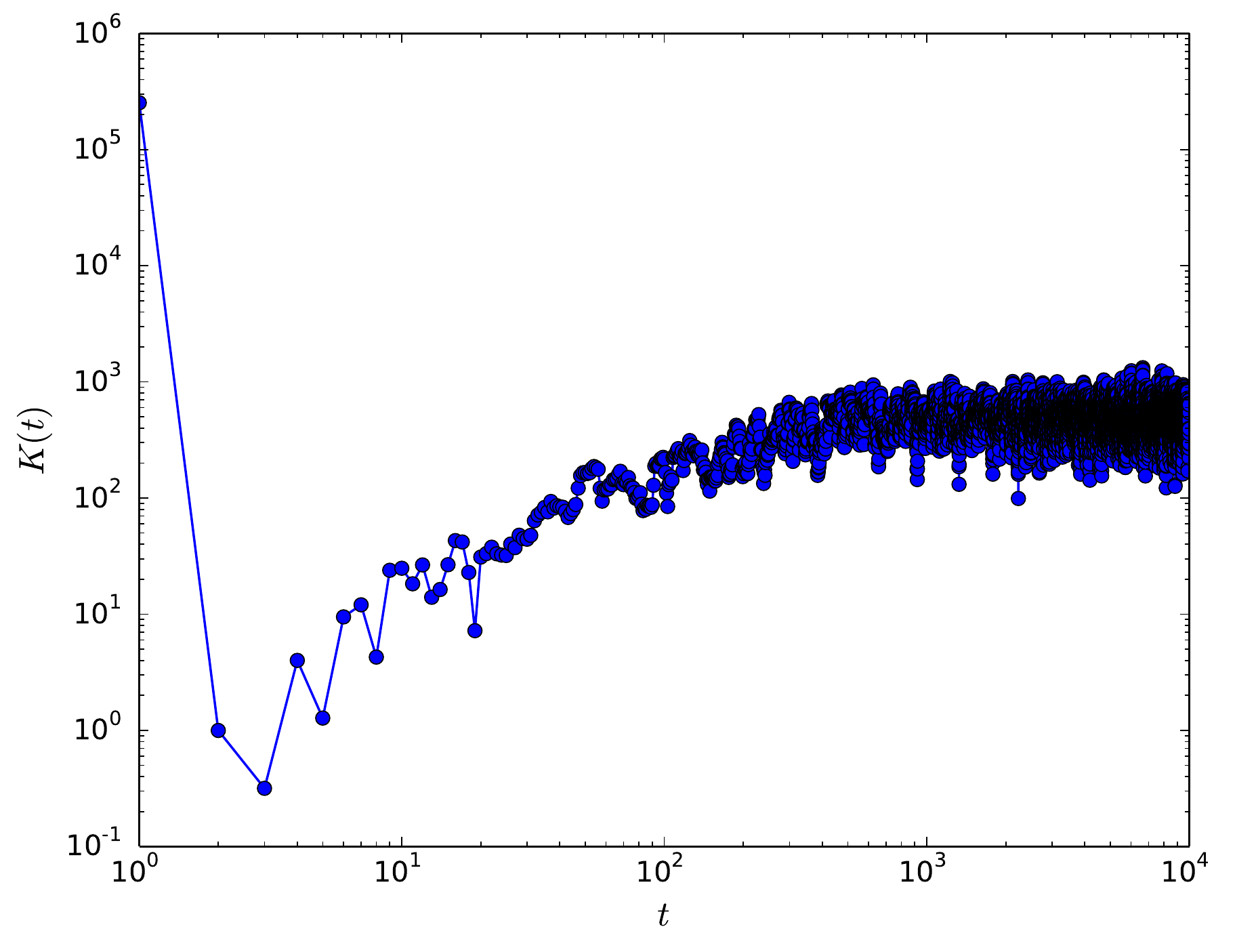}
\caption{The smoothened spectral form factor $K(t)$ for the perturbed cat map with $\kappa = 0.1$ and $N = 503$. Typically the dip time $t_{\textrm{dip}}$ appears to be of the order of operator scrambling time $t_{\textrm{scr}}$ and the plateau time $t_{\textrm{plat}}$ of the order of $N t_{\textrm{scr}}$.}
\label{fig:formfactor}
\end{figure}
We now relate the operator evolution described in Sec.~\ref{sec:regimes} to the spectral form factor, a widely used diagnostic of quantum chaos, further elucidating the basis-independence of our results.
The spectral form factor $K(t)$ is defined as\cite{brezin1997spectral, heusler2004universal, cotler2017black}
\begin{equation}
    K(t) \equiv |\textrm{Tr}(U^t)|^2 = \sumal{m,n}{}{e^{i\left(\phi_m - \phi_n\right)t}},
\label{eq:formfactor}
\end{equation}
where $\{\phi_m\}$ are the quasi-energies of the unitary matrix of the quantum map.
In generic non-integrable systems, $K(t)$ is believed to show three distinct features: a dip, a ramp and a plateau. \cite{cotler2017black, cotler2017chaos} These features can be seen numerically for several systems\cite{cotler2017black} and also in cases where $K(t)$ can be analytically computed. \cite{cotler2017chaos, bertini2018exact}
When $U$ is an $N \times N$ CUE random matrix, $K(t)$ assumes the following values\cite{mehta2004random} 
\begin{equation}
    K(t) = \threepartdef{N^2}{t = 0}{|t|}{0 < |t| \leq N}{N}{|t| \geq N}.
\label{eq:KtCUE}
\end{equation}
To relate operator evolution to the spectral form factor, we note that after an operator basis transformation $K(t)$ of Eq.~(\ref{eq:formfactor}) can be written as 
\begin{eqnarray}
    K(t) &=& \sumal{i=1}{N^2}{\textrm{Tr}\left(\hat{O}_i(0)^\dagger \left(U^\dagger \right)^t \hat{O}_i(0) U^t\right) } \nn \\
    &=& \sumal{i = 1}{N^2}{\textrm{Tr}\left(\hat{O}_i^\dagger(0) \hat{O}_i(t)\right)}, 
\end{eqnarray}
where $\{\hat{O}_i(0)\}$ is a complete orthonormal basis of operators at $t = 0$ and $\{\hat{O}_i(t)\}$ are the time-evolved basis operators.
If $\hat{O}_i(t)$ is expressed in the basis of $\{\hat{O}_i(0)\}$ as
\begin{equation}
    \hat{O}_i(t) = \sumal{i = 1}{N^2}{g_{ij}(t) \hat{O}_j(0)}, 
\label{eq:genbasisexpansion}
\end{equation}
then 
\begin{equation}
    K(t) = \sumal{i = 1}{N^2}{g_{ii}(t)}.
\end{equation}
In the previous section, we studied the evolution for the operator coefficients in the Fourier basis \{$X^m Z^n$\}. 
Choosing $\{O_i(0)\}$ to be the Fourier basis in Eq.~(\ref{eq:genbasisexpansion}), we obtain
\begin{equation}
    K(t) = \sumal{m,n=1}{N}{g_{m,n;m,n}(t)}, 
\label{eq:Kt}
\end{equation}
where 
\begin{equation}
    g_{m,n;m,n}(t) \equiv \frac{1}{N}\textrm{Tr}\left(Z^{-n}X^{-m} X(t)^m Z(t)^n\right),
\label{eq:gmnmn}
\end{equation}
the operator coefficient of $X(t)^m Z(t)^n$ corresponding to the basis element $X^m Z^n$.
The behavior of $g_{m,n;m,n}(t)$ of Eq.~(\ref{eq:gmnmn}) can be deduced using the evolution of operator coefficients of the initial operator $X^m Z^n$, and its overlap with $X^m Z^n$.
At $t = 0$, $K(t) = N^2$ since $g_{m,n;m,n}(0) = 1$.
At early-times, $g_{m,n;m,n}(t)$ decreases as the operator coefficients of $X(t)^m Z(t)^n$ move away from $(m, n)$, as illustrated in Fig.~\ref{fig:phasespace}b.
In Eq.~(\ref{eq:Kt}), if $g_{m,n;m,n}(t)$ has a magnitude of $1/N$ and a random phase, we obtain $K(t)$ as a sum of $N^2$ terms with random phases and magnitudes of $\mathcal{O}(1/N)$, thus $K(t) \sim \mathcal{O}(1)$.
This time, when $K(t) \sim \mathcal{O}(1)$ is known as the dip-time in literature.\cite{cotler2017black, cotler2017chaos} 
In the perturbed cat map, $g_{m,n;m,n}(t)$ has a magnitude of $1/N$ first when $t \sim t_{\textrm{scr}}$, when all of the operator coefficients have a magnitude $\mathcal{O}(1/N)$ since the operator entropy $S_Q(t)$ saturates to $S_Q(t) \approx 2 \log N$.
Thus we expect
\begin{equation}
    t_{\textrm{dip}} \sim t_{\textrm{scr}},
\label{eq:diptime}
\end{equation}
which we also observe numerically, e.g. in Fig.~\ref{fig:formfactor}. It is thus reasonable to assume that evolution by $U^{t_\textrm{scr}}$ is equivalent to evolution by a single time-step with a Haar random matrix. 
At times greater than the dip time, $K(t)$ exhibits a linear increase on average, characteristic of evolution by a random matrix.\cite{kos2017many} 
Furthermore, at timescales much larger than the inverse smallest spacing of the quasi-energy spectrum, any $e^{i \left(\phi_m - \phi_n\right) t}$ is a random phase, and hence $K(t) \sim \mathcal{O}(N)$ due to the $N^2$ random phases in Eq.~(\ref{eq:formfactor}).
Thus, in the operator language, the phases of the $N^2$ terms in Eq.~(\ref{eq:Kt}) are correlated at late times, although they have a magnitude of $\mathcal{O}(1/N)$.  
In the perturbed cat map, since the evolution by $U^{t_{\textrm{scr}}}$ is equivalent to random matrix evolution by one time step, the plateau time can be estimated to be $N$\cite{chan2017solution} in units of the operator scrambiling time,
\begin{equation}
t_{\textrm{plat}} \sim N t_{\textrm{scr}}. 
\label{eq:plateautime}
\end{equation}
This timescale is also observed in Fig.~\ref{fig:formfactor}.
We believe that Eq.~(\ref{eq:plateautime}) is the correct scaling of the plateau time as opposed to $N$ (the inverse of the naive estimate of the smallest spacing) because we numerically find that $t_{\textrm{plat}} \rightarrow \infty$ as $\kappa \rightarrow 0$ ($t_{\textrm{scr}} \rightarrow \infty$).
Note that while we observe the timescales of Eqs.~(\ref{eq:diptime}) and (\ref{eq:plateautime}) to typically hold numerically in smoothened plots of $K(t)$ (e.g. Fig.~\ref{fig:formfactor}), their precise physical interpretation is unclear.  
Firstly, since the spectral form-factor is not self-averaging,\cite{prange1997spectral} and we have a single unitary $U$, rigorous definitions of the dip and plateau times are not clear. 
Furthermore, since we have a single unitary $U$ corresponding to a cat map, how does one define the randomness of $U^t$?
Moreover, even within random matrix theory, notions of randomness for $N \times N$ random matrices are defined in the $N \rightarrow \infty$ limit.
In the perturbed cat map, this limit corresponds to the semi-classical limit, further clouding the definition of randomness in the quantum problem. 
We note that our analysis in Secs.~\ref{sec:regimes} and \ref{sec:spectrum} has some overlap with the recent work of Ref.~[\onlinecite{chen2018operator}], where the perturbed cat map was numerically studied using several diagnostics. 
Furthermore, the Ehrenfest time and the operator scrambling time we have obtained are related to timescales that appear in the early time decay and long time saturation of OTOCs in the perturbed cat map studied in Ref.~[\onlinecite{garcia2018chaos}].
Thus, as shown in Ref.~[\onlinecite{garcia2018chaos}], the operator scrambling time $t_{\textrm{scr}}$ can perhaps be related to Ruelle-Pollicott resonances,\cite{haake2013quantum} determined by the spectrum of the Koopman operator $\bM_C$ (see Eqs.~(\ref{eq:classicalmatrix}) and (\ref{eq:koopmanmatrixel})).
\section{Regular Systems and Mixed Phase Spaces}\label{sec:quantumscars}
\begin{figure}[tb]
\includegraphics[scale=0.47]{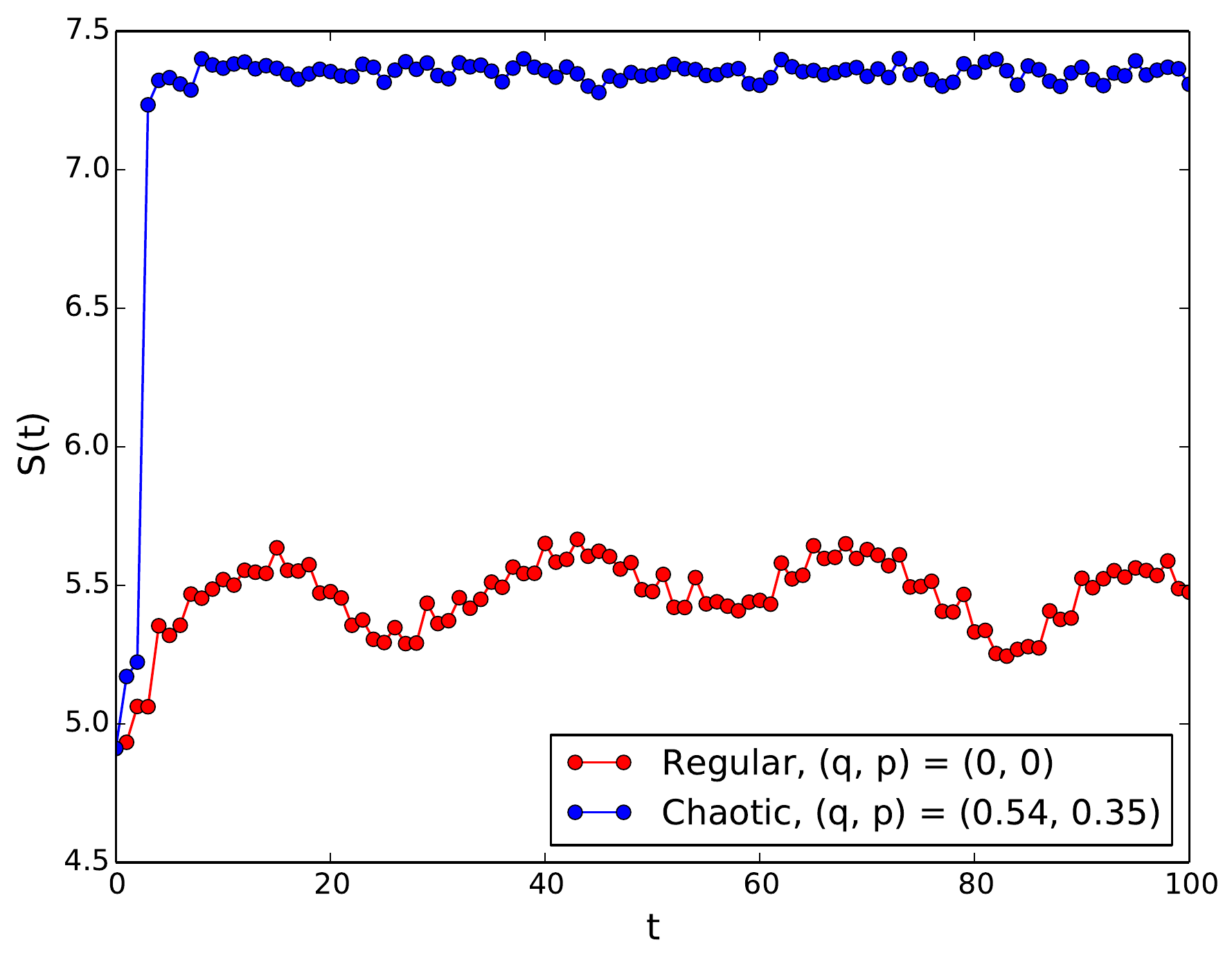}
\caption{(Color online) The quantum entropies of the evolution of a ``coherent operator" $\ket{q\spa p}\bra{q\spa p}$ in a Chirikov Standard map for $\kappa = 3.0$ and $N = 50$. The entropy for a coherent operator in a regular region oscillates around a lower entropy than the one in a chaotic region.}
\label{fig:coherent}
\end{figure}
\begin{figure*}[ht!]
\centering
 \begin{tabular}{cc}
\includegraphics[scale = 0.5]{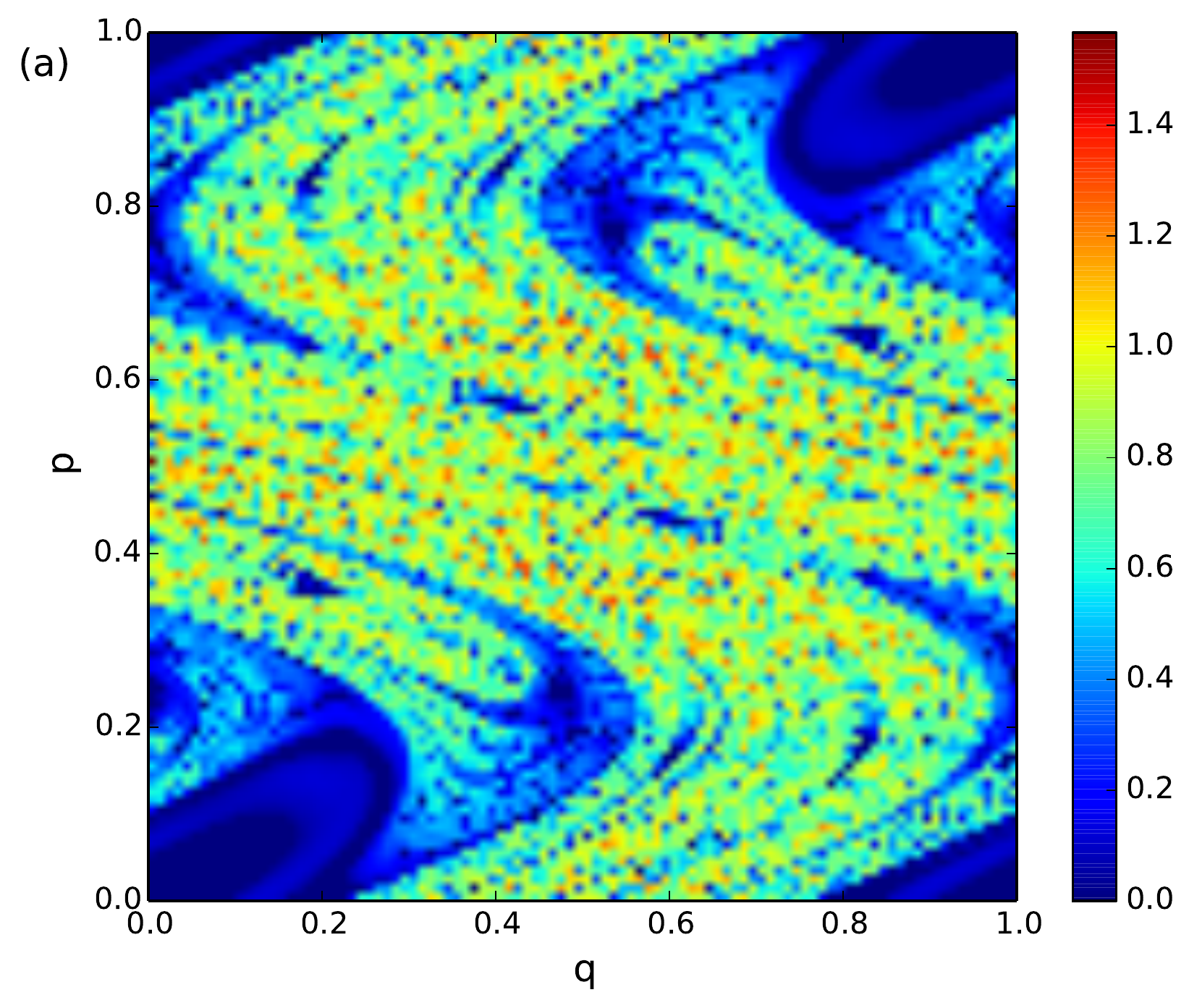}&\includegraphics[scale = 0.5]{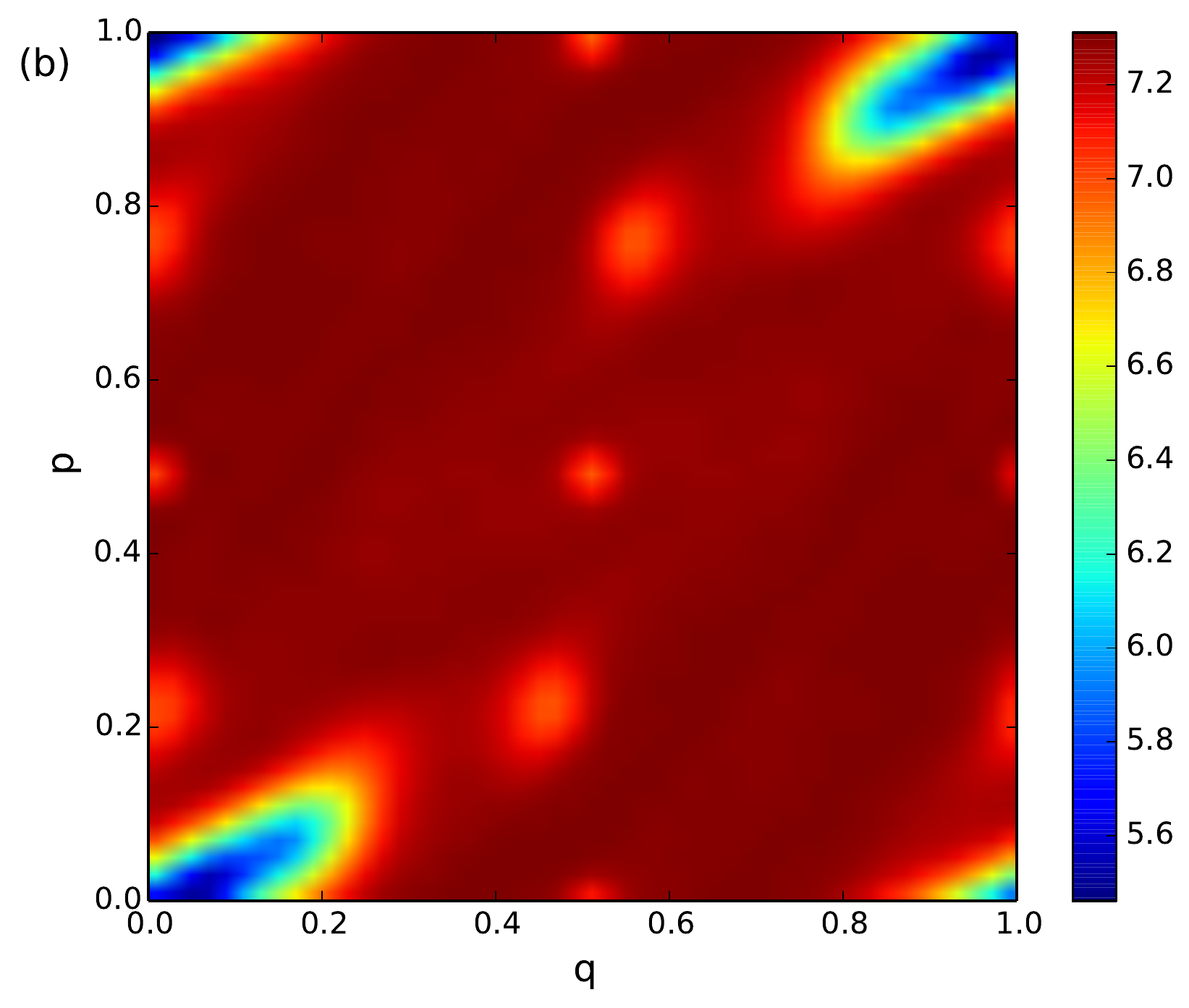}
\end{tabular}
\caption{(Color online) (a) Positive real part of the classical Lyapunov exponents at various points in phase space for the classical Chirikov Standard Map at $\kappa = 3.0$. The blue regions indicate the existence of regular islands. (b) Quantum operator entropy for $N = 50$, $\kappa = 3.0$ for the quantized Chirikov Standard Map. }
\label{fig:lyapunovislands}
\end{figure*}
To illustrate some difference with non-chaotic systems, we now study a map very similar to the one of Eq.~(\ref{eq:genpertcatmap}).
Commonly known as the Chirikov standard map,\cite{chirikov1984correlation} the classical map reads
\begin{eqnarray}
    &\begin{pmatrix}
        q' \\
        p'
    \end{pmatrix}
    =
    \begin{pmatrix}
        1 & 1 \\
        0 & 1
    \end{pmatrix}
    \begin{pmatrix}
        q \\
        p
    \end{pmatrix}\nn \\
    &+
    \frac{\kappa}{2\pi}\sin(2\pi q)
    \begin{pmatrix}
        1 \\
        1
    \end{pmatrix}
    {\rm mod\spa 1}
\label{eq:genstandardmap}
\end{eqnarray}
This map has a zero Lyapunov exponent at $\kappa = 0$ and is known to be non-chaotic for small $\kappa < \kappa_c \approx 1$.\cite{chirikov1984correlation, chirikov2008chirikov, backer2003numerical} 
An interesting feature of the Chirikov standard map is that for a certain range of $\kappa$ the phase space is mixed, i.e. it has both regular and chaotic regions that co-exist in different parts of phase space. \cite{backer2003numerical, lock2010regular}
In terms of the natural variables on a torus (see Eq.~(\ref{eq:compactvars})), the standard map reads
\begin{eqnarray}
    &&z^\prime = x z \exp\left({-\frac{\kappa}{2}(z - z^{\mm})}\right) \nn \\
    &&x^\prime = x \exp\left({-\frac{\kappa}{2}(z - z^{\mm})}\right)
\label{eq:classicalstandardmapcompact}
\end{eqnarray}
Similar to the Arnold Cat map, the classical and quantum evolutions of this map can be compared in the Heisenberg picture via the Fourier and operator coefficients. 
In contrast to the perturbed cat map,the classical entropy in the Chirikov standard map does not show a linear growth for small $\kappa$. 
This is consistent with the fact that the standard map is not chaotic at small $\kappa$ and unlike the cat map (see Eq.~(\ref{eq:comcoefficient})), Fourier coefficients for small $m, n$ do not evolve into ones with exponentially large $m,n$.
Indeed, in the Chirikov Standard Map with $\kappa = 0$, using Eq.~(\ref{eq:standardmapclasscoeffapp}) the coefficients $\{C^{(t)}_{m,n}\}$ after a time $t$ can be related to the initial set of coefficients $\{C^{(0)}_{m,n}\}$ according to 
\begin{equation}
    C^{(0)}_{m,n} = C^{(t)}_{m + n t, n}.
\label{eq:comcoefficient2}
\end{equation}
While the standard map does have an Ehrenfest time for small $\kappa$, estimates such as the ones in Eq.~(\ref{eq:ehrenfesttime}) are no longer accurate, presumably due to the presence of ``hidden" (almost) conserved quantities that need not have simple forms in the Fourier basis. 
Such quantities cause recurrences in the quantum entropies over small timescales, and their existence is indicated by the fact that the Standard map does not exhibit any level repulsion even when $\kappa > 0$.  
Furthermore, the quantum entropy $S_Q(t)$ (defined in Eq.~(\ref{eq:entropies})) never appears to saturate to its maximum value of $2 \log N$ for any $N$.
Such non-chaotic maps have an operator scrambling time $t_{\textrm{scr}} \rightarrow \infty$, due to the existence of conserved or almost-conserved quantities.
In certain classical maps on a torus, chaotic and regular regions can coexist on the phase space.\cite{strelcyn1991coexistence, backer2003numerical}, forming a so-called mixed phase space.
Such behavior is known to exist for large values of $\kappa$ for the perturbed Cat maps and the Chirikov standard map. For example, for the Chirikov standard map at $\kappa = 3.0$, the Lyapunov exponent at various points in phase space is shown in Fig.~\ref{fig:lyapunovislands}a.  A similar phenomenon occurs for the perturbed cat map at $\kappa = 6.5$.\cite{backer2003numerical}
In the quantized maps, this feature manifests as atypical eigenstates of the time evolution operator. 
Indeed, the atypical eigenstates cause the level statistics of the unitary to deviate from both the chaotic and the Poisson distributions.\cite{berry1984semiclassical}
These can be detected by studying the phase space representations (e.g. Husimi functions $H_n(q, p) \equiv |\braket{q\spa p\spa}{\psi}_n|^2$) of the eigenstates $\{\ket{\psi}_n\}$, which look different for the atypical eigenstates.\cite{backer2003numerical}  
The effect of these regular regions in the classical map also show up as features of the density of states of the quantized system, leading  to so-called ``quantum scars".\cite{bogomolny1988smoothed, berry1989quantum, huang2009relativistic}
Several works have studied the fate of these regular islands in quantized versions of these torus maps. 
For example, it is known that in the long-time limit any localized wavepacket that originates in the ergodic region of the phase space eventually ``floods" the regular regions. \cite{backer2005flooding, lock2010regular, backer2002amplitude}
Here we propose a Heisenberg picture interpretation of regular (non-chaotic) islands that appear in the phase portraits of classical maps on $\mathbb{T}^2$. 
Similar to the previous sections, we study the evolution of operator coefficients of an initial ``coherent-state" operator $\ket{q\spa p}\bra{q\spa p}$ where $(q,p)$ is the location of the regular islands in the classical phase space. 
Interestingly, we find in Fig.~\ref{fig:coherent}, the entropy of the operator coefficients for a coherent state \emph{does not} saturate to the maximum value of $2\log N$ at late-times. 
This is indicative of the fact that even at long-times the operator in the regular region does not look resemble a matrix that looks random in the Fourier basis. 
In contrast, a coherent state in the chaotic region \emph{does} saturate to its maximum value in the long-time limit, showing that operator entropy can be used as a diagnostic to detect scars in the spectrum of quantized maps. 

\section{Conclusions}\label{sec:conclusions}
In this work, we have explored classical and quantum chaos in quantum maps in the Heisenberg picture.
We observed that the evolution of operator coefficients in a fixed basis of operators show signatures of quantum chaos in the system.
We compared the behavior of operator coefficients to the behavior of classical Fourier coefficients of functions on phase space, illustrating the differences between the classical equations of motion and the Heisenberg equations of motions.
We obtained a sharp definition of \emph{classicality} of a system and provided examples in which classicality arises away from the usual classical limit. 
We then identified three regimes in the system that show the transition from the early-time classical chaos to late-time quantum chaos, and they are characterized by the natures of evolution of the operator coefficient entropy $S_Q(t)$, defined in Eq.~(\ref{eq:entropies}).
We found that up to an Ehrenfest time $t_E$, the quantum system mimicks the behavior of the classical system and formed the early-time or semiclassical regime. 
Furthermore, after an operator scrambling time $t_{\textrm{scr}}$, the information of the initial operator is scrambled into all the $N^2$ Fourier coefficients, after which the evolution of the coefficients looks random in the Fourier basis. 
This operator scrambling time is the same as the scrambling time for a coherent-state wavefunction to evolve into a wavefunction that is uniformly spread across the system. 
Finally, we used the operator coefficient diagnostics to obtain an operator interpretation of regular islands in torus maps whose classical limits have a mixed phase space.
Our approach characterizes chaos in quantum maps in the Heisenberg picture, which complements previous approaches using wavefunctions (see Ref.~[\onlinecite{backer2003numerical}] and the references therein). 
While we have used the perturbed cat map as an illustrative example, we believe that several aspects of operator evolution in chaotic maps are universal, e.g. the linear and logarithmic growths of the coefficient entropy. 
Since the Heisenberg picture is the natural language to explore many-body quantum chaos and operator spreading in many-body quantum systems\cite{cotler2017black, nahum2017quantum, nahum2017operator, von2017operator, khemani2017operator, rakovszky2017diffusive, hosur2016chaos} and quantum field theories,\cite{mezei2017entanglement1, mezei2017entanglement, maldacena2016bound} our results extend the notions therein to quantum maps on a torus. 
An important open question is to explore if there is a version of ``operator hydrodynamics" for single-particle quantum chaotic systems. \cite{von2017operator}
Furthermore, the analytic tractability of the Heisenberg equations of motion for these maps could further explore connections to information-theoretic concepts such as unitary designs and the complexity growth of states. \cite{cotler2017chaos, roberts2017chaos} 
On a different note, the results in Sec.~\ref{sec:quantumscars} suggest that it would be interesting to study operator evolution in quantum many-body systems that are thought to exhibit many-body quantum scars. \cite{turner2017quantum, moudgalya2017exact, moudgalya2018entanglement, turner2018quantum, ho2018periodic}
\section*{Acknowlegdements}
We thank Arnd B\"acker for an introduction to quantum maps, Roderich Moessner and Philip Holmes for useful discussions and Arul Lakshminarayanan for answers to 
early queries.  SLS is supported by US Department of Energy grant No. DE-SC0016244
\appendix
\onecolumngrid
\section{Heisenberg Equations of Motion}\label{app:heisenbergeqnderivation}
In this section, we derive the Heisenberg equations of motion for the perturbed Cat map and the Chirikov standar map. 
For now, we assume a general map of the form of Eq.~(\ref{eq:genpertcatmap}), and we will restrict ourselves to special values of $a, b, c$ and $d$ when required.
We first represent the $X$ and $Z$ operators (defined in Eq.~(\ref{eq:XZproperties})) as $N \times N$ matrices whose elements read (the indices are represented using $z \equiv \omega^j$ and $z' \equiv \omega^{j'}$)
\begin{equation}
    Z_{z', z} = z \delta_{z', z}, \;\;\; X_{z',z} = \delta_{z', z\omega^{\mm}} 
\label{eq:XZmatrixelements}
\end{equation}
In terms of $z$ and $z'$, by using the substitutions $j = \frac{N }{2\pi i}\log z$, $j' = \frac{N}{2\pi i}\log z'$, the unitary of Eq.~(\ref{eq:unitarycatmap}) can be written as
\begin{eqnarray}
    &(U_N)_{z', z} = \frac{1}{\sqrt{Nb}} \exp\left(\frac{-i N}{4 \pi b} \left(d (\log z)^2 - 2 \log z \log z' + a (\log z')^2 \right)\right. \nn \\
    &\left.+ \frac{\kappa N}{4 \pi}\left(z - z^{\mm}\right)\right)
\label{eq:unitaryzzp}
\end{eqnarray}
$U_N$ can then be written as a product of two matrices, 
\begin{equation}
    (U_N)_{z', z} = \sum_{z''}{U^{(1)}_{z', z''} U^{(2)}_{z'', z}} 
\label{eq:unitarypertunpert}
\end{equation}
where $U^{(1)}$ and $U^{(2)}$ read
\begin{eqnarray}
    &&\hspace{-4mm}U^{(1)}_{z', z''} \equiv \frac{1}{\sqrt{N b}} \exp\left(\frac{-i N}{4 \pi b} \left(d (\log z)^2 - 2 \log z \log z' + a (\log z')^2 \right)\right) \nn \\
    &&U^{(2)}_{z'', z} \equiv \exp\left(\frac{\kappa N}{4 \pi}\left(z - z^{\mm}\right)\right) \delta_{z'', z}
\label{eq:U1U2catmap}
\end{eqnarray}
In Eq.~(\ref{eq:U1U2catmap}), the $U^{(1)}$ corresponds to the unitary for the unperturbed cat map and $U^{(2)}$ corresponds to the perturbation. 
We first expand $U^{(1)}$ in an orthonormal operator basis $\{X^m Z^n\}$ as
\begin{equation}
    U^{(1)} = \sum_{m, n = 0}^{N-1}{u_{m,n} X^m Z^n}.
\label{eq:U1expansion}
\end{equation}
Since the basis is orthonormal, $u_{m,n}$ reads
\begin{equation}
    u_{m,n} = \frac{1}{N}\textrm{Tr }[Z^{-n} X^{-m} U^{(1)}]
\label{eq:umndefinition}
\end{equation}
Thus, we obtain
\begin{eqnarray}
    u_{m,n} &=& \frac{1}{N\sqrt{N b}}\sum_{z'', z'''}{\left(\delta_{z, z''} \delta_{z'', z''' \omega^m} (z'')^{-n} \times \exp\left(\frac{N}{4\pi i b}\left(d (\log z''')^2 - 2 \log z'''\log z + a (\log z)^2\right)\right)\right)} \nn \\
    &=& \frac{1}{N\sqrt{N b}}\sum_{z}{\left(z^{-n} \times \exp\left(\frac{N}{4\pi i b}\left(d (\log z - m \log \omega)^2 - 2 (\log z - m \log \omega) \log z + a (\log z)^2\right)\right)\right)} \nn \\
    &=& \frac{1}{N\sqrt{N b}}\sum_{k = 0}^{N - 1}{\left(\omega^{-kn}\exp\left(\frac{i \pi}{N b}\left((a + d - 2)k^2 - 2 m (d-1) k  + m^2 d\right)\right)\right)} \nn \\
    &=& \frac{1}{N \sqrt{N b}}\sum_{k = 0}^{N -1}{\omega^{\left(\frac{a + d - 2}{2b} k^2 - \left(\frac{m (d -1)}{b} + n\right)k + \frac{m^2 d}{2 b}\right)}}
\label{eq:umngeneral}
\end{eqnarray}

\subsection{Perturbed cat maps}
We first consider a conventional choice of coefficients for the perturbed cat maps where $(a, b, c, d) = (2, 1, 3, 2)$.
The expression for $u_{m,n}$ in Eq.~(\ref{eq:umngeneral}) reduces to 
\begin{equation}
    u_{m,n} = \frac{1}{N \sqrt{N}} \omega^{m^2} \sum_{k = 0}^{N-1}{\omega^{k^2 - (m + n) k}}
\label{eq:umnperturbedcat}
\end{equation}
One might recognize the sum in Eq.~(\ref{eq:umnperturbedcat}) to be a generalized quadratic Gauss sum $G(1, -(m+n), N)$, where 
\begin{equation}
    G(a, b, c) \equiv \sum_{k = 0}^{c - 1}{e^{2\pi i\frac{a n^2 + b n}{c}}}
\end{equation}
Using standard methods for Gaussian sums,\cite{berndt1998gauss} the exact expression for $U^{(1)}$ reads\cite{ hannay1980quantization, keating1991cat}
\begin{equation}
    U^{(1)} = \frac{1 + i}{2 N}\sum_{m, n}{\left( \omega^{\frac{(3m + n)(m - n)}{4}} \left(1 + (-i)^{N + 2 (m + n)}\right) X^m Z^n\right)}
\label{eq:U1pertcatmap}
\end{equation}
Using Eqs.~(\ref{eq:U1U2catmap}) and the matrix elements (\ref{eq:XZmatrixelements}), $U^{(2)}$ can directly be written as
\begin{equation}
    U^{(2)} = \exp\left(\frac{\kappa N}{4 \pi}\left(Z - Z^{\mm}\right)\right)
\label{eq:U2pertcatmap}
\end{equation}
We now proceed to the derivation of the Heisenberg equations of motion. 
By definition, the Heisenberg equations of motion read
\begin{eqnarray}
    &&Z' = {U^{(2)}}^\dagger {U^{(1)}}^\dagger Z U^{(1)} U^{(2)} \nn \\
    &&X' = {U^{(2)}}^\dagger {U^{(1)}}^\dagger X U^{(1)} U^{(2)}.
\label{eq:ZpXppertcat}
\end{eqnarray}
We first focus on computing
\begin{eqnarray}
    &&Z^{(1)} \equiv {U^{(1)}}^\dagger Z U^{(1)}.
\end{eqnarray}
Using the expression for $U^{(1)}$ in Eq.~(\ref{eq:U1pertcatmap}), we obtain
\begin{eqnarray}
    Z^{(1)} &=& \frac{1}{2N^2} \sum_{m,m',n,n'}{\left( \omega^{-\frac{(3m + n)(m-n)}{4}} (1 + i^{N + 2(m+n)})(1 + (-i)^{N+ 2(m'+n')}) \omega^{\frac{(3m' + n')(m' - n')}{4}} Z^{-n} X^{-m} Z X^{m'} Z^{n'}\right)} \nn \\
    &=& \frac{1}{2 N^2}\sum_{m, k, n, l}{\left(\omega^{\frac{1}{4}(3 k^2 - (l - 1)^2 - 2 k (l + 1)} \omega^{\frac{3k - l - 1}{2} m} \omega^{\frac{k - l + 1}{2} n} (1 + i^N (-1)^{m+n} + (-i)^N (-1)^{m+k+n+l - 1} + (-1)^{k+l-1}) X^k Z^l\right)} \nn \\
    &=& \frac{1}{2 N^2}\sum_{k, l}{\left(\omega^{\frac{1}{4}(3 k^2 - (l - 1)^2 - 2 k (l + 1)} (1 + i^N + (-i)^N (-1)^{k+l - 1} + (-1)^{k+l-1}) N \delta_{3k - l - 1, 0}N \delta_{k - l + 1, 0}\right)X^k Z^l} \nn \\
    &=& \omega^{\mm} X Z^2
\label{eq:pertcatmapZ1}
\end{eqnarray}
where in the second line we have defined $k \equiv m' - m$ and $l \equiv n' - n + 1$ and in the third line we have evaluated the sums over $m$ and $n$. 
Similarly, the expression for $X^{(1)}$ can be written as
\begin{eqnarray}
    X^{(1)} &=& \frac{1}{2N^2} \sum_{m,m',n,n'}{\left( \omega^{-\frac{(3m + n)(m-n)}{4}} (1 + i^{N + 2(m+n)})(1 + (-i)^{N+ 2(m'+n')}) \omega^{\frac{(3m' + n')(m' - n')}{4}} Z^{-n} X^{-m} X X^{m'} Z^{n'}\right)} \nn \\
    &=& \frac{1}{2 N^2}\sum_{m, k, n, l}{\left(\omega^{\frac{1}{4}(k - l - 1)(3 k + l - 3)} \omega^{\frac{3k - l - 3}{2} m} \omega^{\frac{k - l + 1}{2} n} (1 + i^N (-1)^{m+n} + (-i)^N (-1)^{m+k+n+l - 1} + (-1)^{k+l-1}) X^k Z^l\right)} \nn \\
    &=& \frac{1}{2 N^2}\sum_{k, l}{\left(\omega^{\frac{1}{4}(k - l - 1)(3 k + l - 3)} (1 + i^N + (-i)^N (-1)^{k+l - 1} + (-1)^{k+l-1}) N \delta_{3k - l + 3, 0}N \delta_{k - l + 1, 0}\right)X^k Z^l} \nn \\
    &=& \omega^{\text{-}3} X^2 Z^3
\label{eq:pertcatmapX1}
\end{eqnarray}
where in the second line we have defined $k \equiv m' - m + 1$ and $l \equiv n' - n$.
Using Eqs.~(\ref{eq:pertcatmapZ1}), (\ref{eq:ZpXppertcat}) and operator commutation relations of Eq.~(\ref{eq:XZproperties}), we obtain $Z'$ to be (since $Z$ is unitary) 
\begin{eqnarray}
    Z' &=& e^{-\frac{\kappa N}{4\pi} (Z - Z^{\mm})} \omega^{\mm} X Z^2 e^{\frac{\kappa N}{4 \pi} (Z - Z^{\mm})} \nn \\
    &=& \omega^{\mm} X Z^2 \exp\left(\frac{\kappa N}{4 \pi}\left[ (1 - \omega^{\mm}) Z - ( 1 - \omega) Z^{\mm}\right]\right) \nn \\
    &=& \omega^{\mm} X Z^2 \exp\left(\frac{\kappa N}{4 \pi}(\omega^{\frac{1}{2}} - \omega^{\text{-} \frac{1}{2}})(\omega^{\text{-} \frac{1}{2}} Z + \omega^{\frac{1}{2}}Z)\right). \nn \\
\label{eq:Zpheis}
\end{eqnarray}
Similarly, using Eqs.~(\ref{eq:pertcatmapX1}) and (\ref{eq:ZpXppertcat}), we obtain $X'$ to be
\begin{eqnarray}
    X' &=& e^{-\frac{\kappa N}{4\pi} (Z - Z^{\mm})} \omega^{\text{-}3} X^2 Z^3 e^{\frac{\kappa N}{4 \pi} (Z - Z^{\mm})} \nn \\
    &=& \omega^{\text{-}} X^2 Z^3 \exp\left(\frac{\kappa N}{4 \pi}\left[ (1 - \omega^{\text{-}2}) Z - ( 1 - \omega^2) Z^{\mm}\right]\right) \nn \\
    &=& \omega^{\text{-} 3} X^2 Z^3 \exp\left(\frac{\kappa N}{4 \pi}(\omega - \omega^{\mm}) (\omega^{\mm} Z + \omega Z^{\mm})\right) \nn \\
\label{eq:Xpheis}
\end{eqnarray}
Thus, Eqs.~(\ref{eq:Zpheis}) and (\ref{eq:Xpheis}) are the Heisenberg equations of motion for the quantization of the perturbed cat map Eq.~(\ref{eq:genpertcatmap}) with $(a, b, c, d) = (2, 1, 3, 2)$.

\subsection{Chirikov standard map}
Similar to Eq.~(\ref{eq:unitarypertunpert}), the unitary for the standard map can be decomposed into an unperturbed part $U^{(1)}$ and a perturbed part $U^{(2)}$.
Since the unperturbed part of the standard map has the same form as the unperturbed part of the cat map, the expression of $U^{(1)}$ is given by Eqs.~(\ref{eq:U1pertcatmap}) and (\ref{eq:umngeneral}) with $(a, b, c, d) = (1, 1, 0, 1)$.
Thus, we obtain
\begin{equation}
    U^{(1)} = \frac{1}{\sqrt{N}}\sum_{m = 0}^{N-1}{\omega^{m^2/2} X^m}.
\label{eq:standardmapU1}
\end{equation}
The perturbed part $U^{(2)}$ is diagonal in the position basis and is given by
\begin{equation}
    U^{(2)} = \exp\left(\frac{i \kappa N}{4\pi}(Z + Z^{\mm})\right).
\end{equation}
Consequently, the expression for $Z'$ reads 
\begin{equation}
    Z' = \omega^{-\frac{1}{2}} X Z \exp\left({\frac{i \kappa N}{4\pi} (\omega^{\frac{1}{2}} - \omega^{-\frac{1}{2}}) (\omega^{-\frac{1}{2}} Z - \omega^{\frac{1}{2}} Z^{\mm})}\right).
\label{eq:standardmapheisenbergZapp}
\end{equation}
Similarly, the expression for $X'$ reads
\begin{equation}
    X' = X  \exp\left({\frac{i \kappa N}{4\pi} (\omega^{\frac{1}{2}} - \omega^{-\frac{1}{2}}) (\omega^{-\frac{1}{2}} Z - \omega^{\frac{1}{2}} Z^{\mm})}\right).
\label{eq:standardmapheisenbergXapp}
\end{equation}

\section{Evolution of operator coefficients}\label{app:operatorevol}
In this section, we derive the evolution equations of the classical Fourier and quantum operator  coefficients. In particular, we derive an expression for the matrix elements $\bra{m',n'} \bM_C \ket{m,n}$ and $\bra{m',n'} \bM_Q \ket{m,n}$ where the classical and quantum coefficients evolve according to 
\begin{eqnarray}
    C'_{m,n} &=& \sum_{m,n = -\infty}^\infty{\bra{m',n'} \bM_C \ket{m,n} C_{m,n}} \nn \\
    Q'_{m,n} &=& \sum_{m,n = 0}^{N-1}{\bra{m',n'} \bM_Q \ket{m,n} Q_{m,n}}
\end{eqnarray}
respectively.
\subsection{Perturbed cat maps}
We start with the evolution of classical Fourier coefficients. Using the classical evolution of Eq.~(\ref{eq:classicalcatmapcompact}), the evolution of Fourier components is defined using Eq.~(\ref{eq:classicalcoeffevoldefn}).
Substituting Eq.~(\ref{eq:classicalcatmapcompact}) into Eq.~(\ref{eq:classicalcoeffevoldefn}), we obtain 
\begin{eqnarray}
    \sum_{m,n=-\infty}^{\infty}{C_{m,n} {x'}^m {z'}^n} &=& \sum_{m,n}{C_{m,n} x^{2m  + n} z^{3 m + 2 n} \exp\left({\frac{i\kappa}{2}(z + z^{\mm})\left(2m + n\right)}\right)} \nn \\
    &=& \sum_{m,n = -\infty}^{\infty}{\sum_{j = 0}^{\infty}\sum_{r = 0}^j{{C_{m,n} \frac{(i \kappa)^j}{r! (j- r)!} \left(m + \frac{n}{2}\right)^j x^{2m + n} z^{3m + 2n + 2r - j}}}} \nn \\
    &\equiv& \sum_{m',n' = -\infty}^{\infty}{C'_{m', n'} x^{m'} z^{n'}} = \sum_{m',n',m,n = -\infty}^{\infty}{\bra{m', n'} \bM_C \ket{m,n} C_{m, n} x^{m'} z^{n'}} 
\label{eq:cmatrixelements}
\end{eqnarray}
The matrix elements in Eq.~(\ref{eq:cmatrixelements}) read
\begin{eqnarray}
    \bra{m', n'} \bM_C \ket{m,n} &=& \sum_{j = 0}^\infty{\sum_{r = 0}^j{\left[ \frac{(i\kappa)^j }{r! (j -r)!} \left(m + \frac{n}{2}\right)^j \delta_{m', 2m + n} \delta_{n', 3m + 2n + 2r - j}\right]}} \nn \\
    &=& \sum_{s = -\infty}^{\infty}{\sum_{j = |s|}^\infty{\left[ \frac{(i\kappa)^j }{\left(\frac{j + s}{2}\right)! \left(\frac{j - s}{2}\right)!} \left(m + \frac{n}{2}\right)^j \delta_{m', 2m + n} \delta_{n', 3m + 2n + s} \delta_{s + j, \textrm{even}}\right]}} \nn \\
    &=& \sum_{s = -\infty}^\infty{\sum_{l = 0}^\infty{\left[ \frac{(i\kappa)^{2l + |s|} }{\left(l + |s|\right)!\spa l!} \left(m + \frac{n}{2}\right)^{2l + |s|} \delta_{m', 2m + n} \delta_{n', 3m + 2n + s}\right]}} \nn \\
    &=& \sum_{s = -\infty}^\infty{\left[i^{|s|} \mJ_{|s|}\left(\kappa (2m + n)\right) \delta_{m', 2m + n} \delta_{n', 3m + 2n + s} \right]} = \sum_{s = -\infty}^\infty{\left[i^s \mJ_{s}(\kappa(2m + n))\delta_{m', 2m + n} \delta_{n', 3m + 2n + s}\right]}, \nn \\
\label{eq:pertcatmapclasscoeffapp}
\end{eqnarray}
where $\mJ_\nu(x)$ is the $\nu$-th order Bessel function of the first kind. In deriving Eq.~(\ref{eq:pertcatmapclasscoeffapp}), we have used
\begin{equation}
    \mJ_{\nu}(x) = \sumal{l = 0}{\infty}{\frac{(-1)^m}{m! (m + \nu)!}\left(\frac{x}{2}\right)^{2m + \nu}}
\label{eq:besseldefn}
\end{equation}
and the property
\begin{equation}
\mJ_{-\nu}(x) = (-1)^\nu \mJ_{\nu}(x).
\label{eq:besselproperty}
\end{equation}

The derivation in the quantum case is similar, using the Heisenberg equations of motion instead of the classical evolution equations.
Using Heisenberg equations of Eqs.~(\ref{eq:pertcatheisenbergeqs}) and the properties of Eq.~(\ref{eq:XZproperties}), we first obtain 
\begin{eqnarray}
    &&{X'}^m = \omega^{-3m^2} X^{2m} Z^{3m} \exp\left(\frac{\kappa N}{4\pi}(\omega^m - \omega^{\text{-}m})(\omega^{\text{-}m}Z + \omega^m Z^{\mm})\right) \nn \\
    &&{Z'}^n = \omega^{-n^2} X^n Z^{2n} \exp\left(\frac{\kappa N}{4\pi}\left(\omega^{\frac{n}{2}} - \omega^{-\frac{n}{2}}\right)\left(\omega^{\text{-}\frac{n}{2}}Z + \omega^{\frac{n}{2}}Z^{\mm}\right)\right).
\label{eq:XnZmcatmap}
\end{eqnarray}
Consequently, using Eqs.~(\ref{eq:XnZmcatmap}) and (\ref{eq:quantumcoeffevoldefn}), we can write 
\begin{eqnarray}
    \sum_{m,n=0}^{N-1}{Q_{m,n} {X'}^m {Z'}^n} &=& \sum_{m,n = 0}^{N-1}{Q_{m,n} \omega^{-3m^2 - n^2 - 3mn} X^{2m + n} Z^{3m + 2n} \exp\left({\frac{\kappa N}{4 \pi}\left(\omega^{\left(m + \frac{n}{2}\right)} - \omega^{-\left(m + \frac{n}{2}\right)}\right)\left(\omega^{-\left(m + \frac{n}{2}\right)} Z + \omega^{\left(m + \frac{n}{2}\right)} Z^{\mm}\right)}\right)} \nn \\
    &=& \sum_{m,n=0}^{N-1}{Q_{m,n}\omega^{-3m^2 - n^2 - 3mn}\sum_{j= 0}^{\infty}{\sum_{r = 0}^j{\omega^{\left(m + \frac{n}{2}\right)(j - 2r)}\frac{(i \kappa N)^j}{(2\pi)^j r! (j - r)!} \sin^j\left(\frac{\pi}{N}(2m + n)\right)X^{2m + n} Z^{3m + 2n + 2r - j}}}},\nn \\
    &\equiv& \sum_{m',n' = 0}^{N-1}{Q'_{m',n'} X^{m'} Z^{n'}} = \sum_{m',n',m,n = 0}^{N-1}{\bra{m',n'} \bM_Q \ket{m,n} Q_{m,n} X^{m'} Z^{n'}}
\label{eq:qmatrixelements}
\end{eqnarray}
where we have used the properties of Eq.~(\ref{eq:XZproperties}).
To write out the matrix elements in Eq.~(\ref{eq:qmatrixelements}) explicitly, it is useful to define
\begin{equation}
    \delta^{(N)}_{a, b} = \twopartdeft{1}{a = b\ \textrm{mod}\ N}{0}.
\end{equation}
The matrix elements then read 
\begin{eqnarray}
    &&\bra{m',n'} \bM_Q \ket{m,n} = \omega^{-3m^2 - n^2 - 3mn} \sum_{j = 0}^{\infty}{\sum_{r = 0}^j{\left[ \frac{\kappa^j \omega^{\left(m + \frac{n}{2}\right)(j - 2r)}}{r! (j - r)!} \left(\frac{i N}{2 \pi} \sin\left(\frac{\pi}{N}(2m + n)\right)\right)^j  \delta^{(N)}_{m', 2m + n} \delta^{(N)}_{n', 3m + 2n + 2r - j} \right]}} \nn \\
    &&= \omega^{-3m^2 - n^2 - 3mn} \sum_{s = 0}^{N-1}{\sum_{j = 0}^\infty{{\sum_{r = 0}^j{\left[ \frac{(i\kappa)^j \omega^{\left(m + \frac{n}{2}\right)(j - 2r)}}{r! (j - r)!} \left(\frac{N}{2 \pi} \sin\left(\frac{\pi}{N}(2m + n)\right)\right)^j  \delta^{(N)}_{m', 2m + n} \delta^{(N)}_{n', 3m + 2n + s} \delta^{(N)}_{2r - j, s} \right]}}}} \nn \\
    &&= \omega^{-3m^2 - n^2 - 3mn} \sum_{p = -\infty}^{\infty}{\sum_{s = 0}^{N-1}{\sum_{j = 0}^\infty{{\sum_{r = 0}^j{\left[ \frac{(i\kappa)^j }{r! (j - r)!} \left(\frac{N}{2 \pi} \sin\left(\frac{\pi}{N}(2m + n)\right)\right)^j \omega^{\left(m + \frac{n}{2}\right)(j - 2r)} \delta^{(N)}_{m', 2m + n} \delta^{(N)}_{n', 3m + 2n + s} \delta_{2r - j + pN, s} \right]}}}}} \nn \\
    &&=  \omega^{-3m^2 - n^2 - 3mn} \sum_{p = -\infty}^{\infty}{ \sum_{s = 0}^{N-1}{\sum_{j = |p + pN|}^\infty{\left[ \frac{(i\kappa)^j \omega^{\left(m + \frac{n}{2}\right)(pN - s)} }{\left(\frac{j + s + pN}{2}\right)! \left(\frac{j - s- p N}{2}\right)!} \left(\frac{N}{2 \pi} \sin\left(\frac{\pi}{N}(2m + n)\right)\right)^j  \delta_{j + s + p N, \textrm{even}} \delta^{(N)}_{m', 2m + n}\delta^{(N)}_{n', 3m + 2n + s} \right]}}} \nn \\
    &&= \omega^{-3m^2 - n^2 - 3mn} \sum_{p = -\infty}^{\infty}{\sum_{s = 0}^{N-1}{\sum_{l = 0}^\infty{\left[ \frac{(i\kappa)^{2l + |s + p N|} \omega^{\left(m + \frac{n}{2}\right)(pN - s)} }{\left(l + |s + pN|\right)!\ l!} \left(\frac{N}{2 \pi} \sin\left(\frac{\pi}{N}(2m + n)\right)\right)^{2l + |s + pN|}  \delta^{(N)}_{m', 2m + n} \delta^{(N)}_{n', 3m + 2n + s} \right]}}} \nn \\
    &&= \omega^{-3m^2 - n^2 - 3mn} \sum_{p = -\infty}^{\infty}{\sum_{s = 0}^{N-1}{\left[i^{|s + pN|} \mJ_{|s + pN|}\left(\frac{\kappa N}{\pi} \sin\left(\frac{\pi}{N}(2m + n)\right)\right) \omega^{\left(m + \frac{n}{2}\right)(pN - s)}\delta^{(N)}_{m', 2m + n} \delta^{(N)}_{n', 3m + 2n + s}\right]}} \nn \\
    &&= \omega^{-3m^2 - n^2 - 3mn} \sum_{p = -\infty}^{\infty}{\sum_{s = 0}^{N-1}{\left[i^{s + pN} \mJ_{s + pN}\left(\frac{\kappa N}{\pi} \sin\left(\frac{\pi}{N}(2m + n)\right)\right) \omega^{\left(m + \frac{n}{2}\right)(pN - s)}\delta^{(N)}_{m', 2m + n} \delta^{(N)}_{n', 3m + 2n + s}\right]}},
\label{eq:pertcatmapquantcoeffapp}
\end{eqnarray}
where we have used Eqs.~(\ref{eq:besseldefn}) and (\ref{eq:besselproperty}).
Note that in the limit $N \rightarrow \infty$, Eq.~(\ref{eq:pertcatmapquantcoeffapp}) reduces to Eq.~(\ref{eq:pertcatmapclasscoeffapp}).

\subsection{Chirikov standard map}
To obtain the operator coefficient evolution for the Standard map, we follow the same procedure as for the perturbed cat map. 
The classical coefficient evolution matrix elements are obtained using Eqs.~(\ref{eq:classicalcoeffevoldefn}) and (\ref{eq:classicalstandardmapcompact}).
The result reads
\begin{eqnarray}
    \bra{m',n'} \bM_C \ket{m,n} &=& \sum_{j = 0}^\infty{\sum_{r = 0}^j{{\left[ (-1)^{r}\frac{\kappa^j }{r! (j - r)!} \left(\frac{m + n}{2}\right)^j \delta_{m', m + n} \delta_{n', n + 2r - j}\right]}}} \nn \\
    &=& \sum_{s = -\infty}^\infty{\left[(-1)^{s}\mJ_{|s|}(\kappa(m + n)) \delta_{m', m + n} \delta_{n', n + s}\right]}.
\label{eq:standardmapclasscoeffapp}
\end{eqnarray}
Similarly, to determine the quantum evolution equation, using the Heisenberg equations of motion in Eqs.~(\ref{eq:standardmapheisenbergZapp}) and (\ref{eq:standardmapheisenbergXapp}), and the properties of Eq.~(\ref{eq:XZproperties}) we first obtain
\begin{eqnarray}
    &&{X'}^m = X^{m} \exp\left(\frac{i \kappa N}{4\pi}\left(\omega^{\frac{m}{2}} - \omega^{\text{-}\frac{m}{2}}\right)\left(\omega^{\text{-}\frac{(m-1)}{2}}Z + \omega^{\frac{(m-1)}{2}} Z^{\mm}\right)\right) \nn \\
    &&{Z'}^n = \omega^{-\frac{n^2}{2}} X^n Z^{n} \exp\left(\frac{i\kappa N}{4\pi}\left(\omega^{\frac{n}{2}} - \omega^{-\frac{n}{2}}\right)\left(\omega^{\text{-}\frac{n}{2}}Z + \omega^{\frac{n}{2}}Z^{\mm}\right)\right).
\label{eq:XnZmstandardmap}
\end{eqnarray}
Consequently, the quantum evolution matrix elements for the Standard map read
\begin{eqnarray}
    \bra{m',n'} \bM_Q \ket{m,n} &=& \omega^{-\frac{n^2}{2}} \sum_{j = 0}^\infty{\sum_{r = 0}^j{\left[(-1)^r \frac{\kappa^j \omega^{\left(\frac{m + n}{2}\right)(j - 2r)}}{r! (j - r)!} \left(\frac{N}{2 \pi} \sin \left(\frac{\pi}{N}(m + n)\right)\right)^j  \delta_{m', m + n} \delta_{n', n + 2j - r}\right]}} \nn \\
    &=&\omega^{-\frac{n^2}{2}}\sum_{s = 0}^{N-1}{\sum_{p = -\infty}^\infty{\left[(-1)^{s + pN} \mJ_{|s + pN|}\left(\frac{\kappa N}{\pi}\sin\left(\frac{\pi}{N}(m + n)\right)\right) \omega^{\left(\frac{m + n}{2}\right)(pN - s)} \delta_{m', m + n} \delta_{n', n + s} \right]}}. \nn \\
\label{eq:standardmapquantcoeffapp}
\end{eqnarray}
In the limit $N \rightarrow \infty$, Eq.~(\ref{eq:standardmapquantcoeffapp}) reduces to Eq.~(\ref{eq:standardmapclasscoeffapp}).
\section{Bessel Function Approximations}\label{app:spreading}
In this section, we estimate a ``decay length" $\nu_0$ such that the magnitude of the Bessel function $|\mathcal{J}_\nu(x)|$ can be considered to vanish for $|\nu| > \nu_0$. We consider the $|x| \ll 1$ and $|x| \gg 1$ cases separately:
\begin{enumerate}
\item When $|x| \ll 1$, we use the usual expansion of the Bessel functions as
\begin{eqnarray}
    |\mathcal{J}_\nu(x)| &=& \sum_{s = 0}^\infty{\frac{(-1)^s}{\Gamma(s+1)\Gamma(\nu-s+1)}\left(\frac{|x|}{2}\right)^{n+2s}} \approx \frac{1}{\Gamma(\nu+1)}\left(\frac{|x|}{2}\right)^\nu + \mathcal{O}(|x|)  \nn \\
    &\approx& \frac{1}{\Gamma\left(\nu + 1\right)}\exp(\nu \log(|x|/2)) 
\label{eq:besselexpansion}
\end{eqnarray}
Consequently, $|\mathcal{J}_\nu(x)|$ can be considered to vanish for $|\nu| \gtrapprox \nu_0$, where 
\begin{eqnarray}
    \nu_0 = \frac{1}{\log(2/|x|)}.
\label{eq:nu0smallx}
\end{eqnarray}
\item When $|x| \gg 1$, we expect $\nu_0 \gg 1$, and hence we can use the two forms of Debye expansions for Bessel functions\cite{olver1954asymptotic}: 
\begin{eqnarray}
    |\mathcal{J}_\nu\left(\nu \sech \alpha\right)| &\approx& \frac{\exp\left(-\nu (\alpha - \tanh{\alpha})\right)}{\sqrt{2\pi\nu \tanh{\alpha}}}\left(1 + \mathcal{O}\left(\frac{1}{\nu}\right)\right),\;\;\; \nu \rightarrow \infty,\;\;\; \alpha > 0 \label{eq:debyeexpansion1} \\
    |\mathcal{J}_\nu\left(\nu \sec \beta\right)| &\approx& \sqrt{\frac{2}{\pi \nu \tan\beta}}\left(\cos\left(\nu \left(\tan \beta - \beta\right) - \frac{\pi}{4}\right) + \mathcal{O}\left(\frac{1}{\nu}\right)\right),\;\;\; \nu \rightarrow \infty,\;\;\; \beta > 0
\label{eq:debyeexpansion2}
\end{eqnarray}
Substituting $x = \nu \sech \alpha$ and $x = \nu \sec \beta$ in Eqs.~(\ref{eq:debyeexpansion1}) and (\ref{eq:debyeexpansion2}) respectively, and using $\tanh\left(\arcsech y\right) = \sqrt{1 - y^2}$ and $\tan\left(\arcsec y\right) = \sqrt{y^2 - 1}$, we obtain 
\begin{equation}
    |\mathcal{J}_{\nu}\left(x\right)| \approx \twopartdef{\frac{\exp\left(-\nu \arcsech\left(\frac{x}{\nu}\right) + \sqrt{\nu^2 - x^2}\right)}{\sqrt{2\pi \sqrt{\nu^2 - x^2}}}\left(1 + \mathcal{O}\left(\frac{1}{\nu}\right)\right)}{|\nu| > |x|}{\sqrt{\frac{2}{\pi \sqrt{x^2 - \nu^2}}}\left(\cos\left(\sqrt{x^2 - \nu^2} - \nu \arcsec\left(\frac{x}{\nu}\right) - \frac{\pi}{4}\right) + \mathcal{O}\left(\frac{1}{\nu}\right)\right)}{|\nu| < |x|}. \label{eq:deb}
\end{equation}
Thus, using Eq.~(\ref{eq:deb}) we see that $|\mJ_\nu(x)|$ oscillates for $|\nu| < |x|$ and decays for $|\nu| > |x|$. Thus, the ``decay length" for $|x| \gg 1$ can be defined as
\begin{equation}
    \nu_0 = [|x|],
\label{eq:nu0largex}
\end{equation}
where $[x]$ is the integer part of $x$.
\end{enumerate}

\twocolumngrid
\bibliography{heisenberg_chaos}

\end{document}